\documentclass[12pt]{article}
\usepackage{graphicx}

\def\gtap{\raisebox{-.4ex}{\rlap{$\sim$}} \raisebox{.4ex}{$>$}}
\def\xwl {$x_W^l$}

\def\journal{\topmargin 0.0in   \oddsidemargin 0in
        \headheight 0pt \headsep 0pt
        \textwidth 6.5in 
\textheight 9in 
        \marginparwidth 1.5in
        \parindent 2em
        \parskip .5ex plus .1ex         \jot = 1.5ex}
\journal

\begin{document}
\begin{titlepage}

\noindent June 24, 2002      \hfill    LBNL-50718\\

\begin{center}

\vskip .5in

{\large \bf Electroweak Data and the Higgs Boson Mass: \\
     A Case for New Physics}
\footnote
{This work is supported in part by the Director, Office of Science, Office
of High Energy and Nuclear Physics, Division of High Energy Physics, of the
U.S. Department of Energy under Contract DE-AC03-76SF00098}

\vskip .5in

Michael S. Chanowitz\footnote{Email: chanowitz@lbl.gov}

\vskip .2in

{\em Theoretical Physics Group\\
     Ernest Orlando Lawrence Berkeley National Laboratory\\
     University of California\\
     Berkeley, California 94720}
\end{center}

\vskip .25in

\begin{abstract}

Because of two $3\sigma$ anomalies, the Standard Model (SM) fit of the
precision electroweak data has a poor confidence level, $CL= 0.010$.
Since both anomalies involve challenging systematic issues, it might
appear that the SM could still be valid if the anomalies resulted from
underestimated systematic error. Indeed the $CL$ of the global fit
could then increase to 0.65, but that fit predicts a small Higgs boson
mass, $m_H=43$ GeV, that is only consistent at $CL=0.035$ with the
lower limit, $m_H>114$ GeV, established by direct searches.  The data
then favor new physics whether the anomalous measurements are excluded
from the fit or not, and the Higgs boson mass cannot be predicted
until the new physics is understood. Some measure of statistical
fluctuation would be needed to maintain the validity of the SM, which
is unlikely by broad statistical measures. New physics is favored, but
the SM is not definitively excluded.

\end{abstract}

\end{titlepage}

\renewcommand{\thepage}{\roman{page}}
\setcounter{page}{2}
\mbox{ }

\vskip 1in

\begin{center}
{\bf Disclaimer}
\end{center}

\vskip .2in

\begin{scriptsize}
\begin{quotation}
This document was prepared as an account of work sponsored by the United
States Government. While this document is believed to contain correct
 information, neither the United States Government nor any agency
thereof, nor The Regents of the University of California, nor any of their
employees, makes any warranty, express or implied, or assumes any legal
liability or responsibility for the accuracy, completeness, or usefulness
of any information, apparatus, product, or process disclosed, or represents
that its use would not infringe privately owned rights.  Reference herein
to any specific commercial products process, or service by its trade name,
trademark, manufacturer, or otherwise, does not necessarily constitute or
imply its endorsement, recommendation, or favoring by the United States
Government or any agency thereof, or The Regents of the University of
California.  The views and opinions of authors expressed herein do not
necessarily state or reflect those of the United States Government or any
agency thereof, or The Regents of the University of California.
\end{quotation}
\end{scriptsize}

\vskip 2in

\begin{center}
\begin{small}
{\it Lawrence Berkeley National Laboratory is an equal opportunity employer.}
\end{small}
\end{center}

\newpage

\renewcommand{\thepage}{\arabic{page}}
\setcounter{page}{1}

\noindent{\bf 1. Introduction}

A decade of beautiful experiments at CERN, Fermilab, and SLAC have
provided increasingly precise tests of the Standard Model (SM) of
elementary particle physics. The data is important for two reasons: it
confirms the SM at the level of virtual quantum effects and it probes
the mass scale of the Higgs boson, needed to complete the model and
provide the mechanism of mass generation.  In the usual interpretation
the data is thought to constrain the Higgs boson mass, $m_H$, most
recently with $m_H<195$ GeV\cite{ewwg_02} at confidence level
$CL=95\%$.  At the same time direct searches for the Higgs boson at
LEP II have established a 95\% $CL$ lower limit, $m_H>114$
GeV.\cite{mhlimit}\footnote 
{ 
N.B., the experimental 95\% lower limit
from the direct searches does {\em not} imply a 5\% chance that the
Higgs boson is lighter than 114 GeV; rather it means that if the
mass were actually 114 GeV there would be a 5\% chance for it to
have escaped detection. The likelihood for $m_H < 114$ GeV from the
direct searches is much smaller than 5\%. See for instance the
discussion in section 5 of \cite{mchvr}.  
}  

Recently the agreement of the precision data with the SM has moved
from excellent to poor. For the global fits enumerated below, the 
confidence level has evolved from 0.45 in the Summer of
1998,\cite{ewwg_98} to 0.04 in the Spring of 2001,\cite{ewwg_01} and
then to 0.010 in the current Spring 2002 data.\cite{ewwg_02}\footnote
{
$CL=0.010$ for Spring 2002 is from a fit specified below that uses the
same set of measurements as were included in the quoted 1998 and 2001
fits. Reference \cite{ewwg_02} has a slightly different value
for their all-data fit, $CL=0.017$, because of two recently introduced
measurements, which we do not include as discussed below. Furthermore, 
updating the all-data fit of \cite{ewwg_02} we find, as discussed below, 
that it would now yield $CL=0.009$. 
}
The current low $CL$ is a consequence of two $3\sigma$ anomalies,
together with the evolution of the $W$ boson mass measurement, as
shown below. The $3\sigma$ anomalies are (1) the discrepancy between
the SM determination of $x_W^l = {\rm sin}^2\theta^l_W$, the effective
leptonic weak interaction mixing angle, from three hadronic asymmetry
measurements, $x_W^l[A_H]$, versus its determination from three
leptonic measurements, $x_W^l[A_L]$, and (2) the NuTeV measurement of
charged and neutral current (anti)neutrino-nucleon
scattering,\cite{nutev} quoted as an effective on-shell weak
interaction mixing angle, $x_W^{\rm OS}[\stackrel {(-)}{\nu} N]$.

If either anomaly is genuine, it indicates new physics, the
SM fit is invalidated, and we cannot use the precision data to
constrain the Higgs boson mass until the new physics is
understood. However both anomalous measurements involve subtle
systematic issues, concerning experimental technique and, especially,
nontrivial QCD-based models.  If the systematic
uncertainties were much larger than current estimates, the $CL$ of the
global fit could increase to as much as 0.65, as shown below. It is
then possible to imagine that the SM might still provide a valid
description of the data and a useful constraint on the Higgs boson
mass. 

We will see however that this possibility is unlikely, because of a
contradiction that emerges between the resulting global fit and the
95\% $CL$ lower limit, $m_H > 114$ GeV. The central point is that
{\em the anomalous measurements are the only $m_H$-sensitive
observables that place the Higgs boson mass in the region allowed by
the searches}. All other $m_H$-sensitive observables predict $m_H$ far
below 114 GeV. We find that if the anomalous measurements are
excluded, the confidence level for $m_H>114$ GeV from the global fit
is between 0.030 and 0.035, depending on the method of estimation.

The hypothesis that the anomalies result from systematic error then
also favors new physics, in particular, new physics that would raise
the prediction for $m_H$ into the experimentally allowed region. This
can be accomplished, for example, by new physics whose dominant effect
on the low energy data is on the $W$ and $Z$ vacuum polarizations
(i.e., ``oblique''\cite{pt}), as shown explicitly below. Essentially
any value of $m_H$ is allowed in these fits.

It should be clear that our focus on the possibility of underestimated
systematic error is not based on the belief that it is the most likely
explanation of the data. In fact, the situation is truly puzzling, and
there is no decisive reason to prefer systematic error over new
physics as the explanation of either anomaly. Rather we have
considered the systematic error hypothesis in order to understand
its implications, finding that it also points to new physics.

The SM is then disfavored whether the experimental anomalies are
genuine or not. The viability of the SM fit and the associated
constraint on $m_H$ can only be maintained by invoking some measure of
statistical fluctuation, perhaps in combination with a measure of
increased systematic uncertainty. This is {\it a priori} unlikely by
broad statistical measures discussed below, but it is not impossible.
The conclusion is that the SM is disfavored but not definitively
excluded. A major consequence is that it is important to search for the
Higgs sector over the full range allowed by unitarity,\cite{mcmkg} 
as, fortunately, we will be able  to do at the LHC operating at its 
design luminosity.\cite{ww} 

This paper extends and updates a previously published
report,\cite{mc1} based on the Spring 2001 data
set, which focused exclusively on the $m_H$-sensitive observables. The
present analysis is based on the Spring 2002 data, and considers
$m_H$-sensitive observables as well as global fits of all $Z$-pole
observables. The data has also changed in some respects: the $3\sigma$
NuTeV anomaly is a new development and the discrepancy between the
hadronic and leptonic determinations of $x_W^l$ has diminished from
3.6 to 3.0$\sigma$. However the other $m_H$-sensitive observables are
unchanged, and the present conclusions are consistent with the
previous report.

Since in this work we also consider global fits, we can summarize the
conclusion quantitatively by introducing the combined probability
$$
P_C= CL({\rm Global\ Fit}) \times CL(m_H > 114).  \eqno(1.1)
$$
The internal consistency of the global fit and its consistency with
the search limit are independent constraints, so the combined
likelihood to satisfy both is given by $P_C$. We find that $P_C$ is
roughly independent of whether the three hadronic front-back asymmetry
measurements are included in the fit, although the two factors on the
right hand side of eq. (1.1) vary considerably in the two cases. For
instance, for the global fit to `all' data, we have $CL({\rm Global\
Fit})=0.010$ and $CL(m_H >114)= 0.30$ so that $P_C= 0.010 \times 0.30=
0.0030$.  If the three hadronic asymmetry measurements are omitted we
have instead $CL({\rm Global\ Fit})=0.066$, $CL(m_H >114)= 0.047$, and
$P_C= 0.066 \times 0.047= 0.0031$. The extent of the agreement in this
example is accidental, but the point remains approximately valid: if
the three hadronic asymmetry measurements are omitted, the increase in
the global fit confidence level is approximately compensated by a
corresponding decrease in the confidence level that the fit is
consistent with the direct search limit.

In section 2 we review the data used in the fits, with a discussion of
how it has evolved during the past few years which 
emphasizes the importance of the $W$ mass measurement. In section 3
we briefly discuss the three generic explanations ---
statistics/systematics/new physics --- of the discrepancies in the
global SM fit.  In section 4 we review the methodology of the SM fits
and the choice of observables. In section 5 we present fits of the
data which exhibit the range in $m_H$ preferred by the $m_H$-sensitive
observables, as well as global fits with and without the anomalous
measurements. In these fits we use the $\chi^2$ distribution for the
global fits and the $\Delta\chi^2$ method to assess the consistency of
the fits with the direct search lower limit on $m_H$.  In section 6 we
use a ``Bayesian'' maximimum likelihood method instead of
$\Delta\chi^2$ to estimate the $CL$ for consistency with the direct
searches. Section 7 illustrates the possible effect of new physics in
the oblique approximation. The results are discussed in section 8.

\noindent {\bf 2. The Data}

We consider 13 $Z$-pole observables and in addition the directly
measured values of $m_W$, the $W$ boson mass, $m_t$, the top quark
mass, $\Delta \alpha_5$, the hadronic contribution to the
renormalization of the electromagnetic coupling at the $Z$ pole, and
the NuTeV result. As discussed in section 4, we do not include the $W$
boson width or the Cesium atomic parity violation measurement, which
is the principal reason for the small differences between the global
fits presented here and in \cite{ewwg_02}. These measurements have only 
recently been added to the global fits; they were not included in the
1998 and 2001 fits\cite{ewwg_98,ewwg_01} which we also consider below.  Our
all-data fit is tabulated in table 2.1, with the current preliminary
experimental values from \cite{ewwg_02}. Details of the fitting
procedure are given in section 4.

The central value for $x_W^{\rm OS}[\stackrel {(-)}{\nu} N]$ from the
NuTeV experiment is shown in table 2.1. In our SM fits we include the
small dependence of $x_W^{\rm OS}[\stackrel {(-)}{\nu} N]$ on $m_t$
and $m_H$ given in \cite{nutev}.  Table 2.1 also contains the model
independent NuTeV result\cite{nutev}, given in terms of effective
$Zqq$ couplings, $g_L^2= g_{uL}^2 + g_{dL}^2$ and $g_R^2= g_{uR}^2 +
g_{dR}^2$. They are not included in the SM fits but are used instead
of $x_W^{\rm OS}[\stackrel {(-)}{\nu} N]$ in the new physics fits of
section 7.

The confidence level of the SM fit in table 2.1 is poor, $CL=0.010$,
with $\chi^2/N=27.7/13$. The central value of the Higgs boson mass is
$m_H=94$ GeV. As shown in section 4, our results agree very well with
those of \cite{ewwg_02} when we fit the same set of observables.

The SM fit was excellent in 1998 and has now become poor.  Large
discrepancies occur among the six SM determinations of the effective
leptonic weak interaction mixing angle, \xwl. The three leptonic
measurements, $A_{LR}$, $A_{FB}^l$, and $A_{e,\tau}$ are quite
consistent with one another. They combine with $\chi^2/N = 1.6/2$,
$CL= 0.45$, to yield
$$
x_W^l[A_L]= 0.23113 (21).		\eqno(2.1)
$$
The three hadronic measurements are also mutually consistent and
combine, with $\chi^2/N = 0.03/2$ and $CL= 0.985$, to yield
$$
x_W^l[A_H]= 0.23220 (29).	\eqno(2.2)
$$
But $ x_W^l[A_L]$ and $ x_W^l[A_H]$ differ by 2.99$\sigma$
corresponding to $CL= 0.0028$.  Combining (2.1) and (2.2), the result
for all six measurements is $x_W^l= 0.23149 (17)$.  The very small
$\chi^2$ associated with the three hadronic measurements is either a 
fluctuation or it suggests that the errors are overestimated, in
which case the discrepancy between $x_W^l[A_L]$ and $x_W^l[A_H]$ would
be even greater.

The discrepancy between $x_W^l[A_L]$ and $x_W^l[A_H]$ is driven by the
difference of the two most precise measurements, $A_{LR}$ and
$A_{FB}^b$, which has been a feature of the data since the earliest
days of LEP and SLC.  At present, $x_W^l$ from $A_{LR}$ and $A_{FB}^b$
are respectively 0.23098(26) and 0.23218(31). They differ by
2.97$\sigma$, $CL= 0.0030$, and combine to yield $x_W^l= 0.23151
(20)$.

Combining all 6 measurements directly we find $x_W^l= 0.23149(17)$ as
above, with $\chi^2/N = 10.6/5$ and $CL= 0.06$. Notice that the ratio
of this confidence level to the confidence level, $CL= 0.003$, for 
$x_W^l[A_L]$ versus $x_W^l[A_H]$, 0.06/0.003 = 20, is just the number of
ways that two sets of three can be formed from a collection of 6
objects. If one attaches an {\it a priori} significance to the
leptonic and hadronic subsets, then the appropriate confidence level
is 0.003, from the combination of $x_W^l[A_L]$ and $x_W^l[A_H]$.  If
instead one regards the grouping into $x_W^l[A_L]$ and $x_W^l[A_H]$ as one
of 20 random choices, then 0.06 is the appropriate characterization of
the consistency of the data.\footnote{I thank M. Grunewald for a
discussion.}  In either case the consistency is problematic.

The determination of $x_W^l$ from the hadronic asymmetries assumes
that the hadronic $Zq\overline q$ interaction vertices are given by
the SM. For instance, to obtain $x_W^l$ from 
$$
A_{FB}^b = {3\over4} A_b A_e \eqno{(2.3)}
$$ 
we assume that $A_b$ is at its SM value, $A_b = A_b[{\rm
SM}]$. $A_b[{\rm SM}]$ has very little sensitivity to the unknown
value of $m_H$, and not much sensitivity to the other SM parameters
either. $x_W^l$ is then obtained from $A_e = (g_{eL}^2 -
g_{eR}^2)/(g_{eL}^2 + g_{eR}^2)$, using $g_{eL} = {-1\over2} + x_W^l$
and $g_{eR} = x_W^l$. The only assumption in obtaining $x_W^l$ from
the leptonic asymmetries is lepton flavor universality.

The $3\sigma$ discrepancy between $x_W^l[A_L]$ and $x_W^l[A_H]$ is
significant for three reasons. First, it is a failed test for the SM,
since it implies $A_q \neq A_q[{\rm SM}]$.  For instance,
$A_b$ extracted from $A_{FB}^b$ (taking $A_l$ from the three leptonic
asymmetry measurements) disagrees with $A_b[{\rm SM}]$ by 2.9$\sigma$,
$CL= 0.004$.  Second, together with the $m_W$ measurement, the
$x_W^l[A_L]$ -- $x_W^l[A_H]$ discrepancy marginalizes the global SM fit,
even without the NuTeV result. Finally, in addition to the effect on
the global fit, it is problematic that the determination of the Higgs
boson mass is dominated by the low probability combination of
$x_W^l[A_L]$ and $x_W^l[A_H]$, or by the low probablility combination of
the six asymmetry measurements. In judging the reliability 
of the prediction for $m_H$ we are concerned not only with the 
quality of the global fit but also with the consistency of the 
smaller set of measurements that dominate the $m_H$ prediction. 

To understand the effect on the global fit it is useful to consider
the evolution of the data from 1998\cite{ewwg_98} to the
present\cite{ewwg_02}, shown in table 2.2, together with the
intervening Spring '01 data set\cite{ewwg_01}, on which \cite{mc1} was
based.  The $x_W^l[A_L]$ -- $x_W^l[A_H]$ discrepancy evolved from
2.4$\sigma$ in '98 to 3.6$\sigma$ in Spring '01 to 3.0$\sigma$ in Spring
'02. Excluding NuTeV, the $CL$ of the set of measurements listed in
table 2.1 evolved during that time from a robust 0.46 to 0.04 to
0.10.\footnote 
{ 
The degrees of freedom decrease from 14 to 12 because
we follow the recent practice of the EWWG\cite{ewwg_02} in
consolidating the LEP II and FNAL measurements into a single $m_W$
measurement and the two $\tau$ polarization measurements into a single
quantity that we denote $A_{e,\tau}$.  The same set of measurements is
tracked for all three years.  
}

The decrease in the global $CL$ is only partially due to the changes
in the asymmetry measurements. An equally important factor is the
evolution of $m_W$, for which the precision improved dramatically, by
a factor of 3, while the central value increased by ${1\over 2}\sigma$ with
respect to the '98 measurement. To understand the role of $m_W$, table
2.3 shows fits based on the current data plus two hypothetical
scenarios in which all measurements are kept at their Spring '02
values except $m_W$. In the first of these, $m_W$ is held at its '98
central value and precision. In the second the current precision is
assumed but with a smaller central value, corresponding to a
${1\over 2}\sigma$ downward fluctuation of the '98 measurement. 
For both hypothetical data sets, the global $CL$ is greater by
a factor two than the $CL$ of the current data. 

To understand how $m_W$ correlates with the asymmetry measurements we
also exhibit the corresponding fits in which either $A_{FB}^b$ or $A_{LR}$
are excluded. In the current data, the $CL$ increases appreciably, to
0.51, if $A_{FB}^b$ is excluded but much less if $A_{LR}$ is excluded,
reflecting the larger pull of $A_{FB}^b$ in the SM fit.  In the two
hypothetical scenarios the $CL$ increases comparably whether $A_{LR}$ 
or $A_{FB}^b$ is excluded. 

There are two conclusions from this exercise. First, the evolution of
the $m_W$ measurement contributes as much to the marginalization of
the global fit as does the evolution of the asymmetry measurements.
Second, at its current value and precision $m_W$ tilts the SM fit
toward $A_{LR}$ and $x_W^l[A_L]$, while tagging $A_{FB}^b$ and
$x_W^l[A_H]$ as `anomalous'.  The reason for this ``alliance'' of
$m_W$ and $x_W^l[A_L]$ will become clear in section 5, where we will
see that $m_W$ and $x_W^l[A_L]$ favor very light values of the Higgs
boson mass, far below the 114 GeV lower limit, while $x_W^l[A_H]$
favors much heavier values, far above 114 GeV.

Returning to table 2.2, we also see the effect on the global fit of
the new result from NuTeV. In the '98 and '01 data sets, NuTeV had
little effect on the global $CL$. In the current data set, because of
its increased precision and central value, it causes the $CL$ to
decrease from an already marginal 0.10 to a poor 0.010. The low
confidence level of the global SM fit is then due in roughly equal
parts to 1) the discrepancy between the $x_W^l[A_L]$--$m_W$ alliance
versus $x_W^l[A_H]$, and 2) the NuTeV result. We will refer to
$x_W^l[A_H]$ and $x_W^{\rm OS}[\stackrel {(-)}{\nu} N]$ as ``anomalous''
simply as a shorthand indication of their deviation from the SM fit,
with no judgement intended as to their {\it bona fides}.

\noindent {\bf 3. Interpreting the Discrepancies}

In this section we wish to set the context for the fits to follow 
by briefly discussing the three generic explanations of the 
discrepancies in the SM fit reviewed in the previous section. 
They are statistical fluctuation, new physics, and underestimated 
systematic error. Combinations of the three generic options are also 
possible.

\noindent \underline{{\em 3.1 Statistical Fluctuations} }

One or both anomalies could be the result of statistical
fluctuations. However, if the data is to be consistent not only with
the global fit but also with the lower limit on $m_H$ from the Higgs
boson searches as discussed in sections 5 and 6, it is necessary that
both anomalous and non-anomalous measurements have fluctuated. If only
the anomalous measurements were to have fluctuated, the global fit
would improve but the conflict with the lower limit on $m_H$ would be
exacerbated.

A high energy physics sage is reputed to have said, only partly in
jest, that ``The confidence level for $3\sigma$ is fifty-fifty.''
The wisdom of the remark has its basis in two different
phenomena. First, at a rate above chance expectation, many unusual
results are ultimately understood to result from systematic
error --- this possibility is discussed below and its
implications are explored in sections 5 and 6. Second, estimates of
statistical significance are sometimes not appropriately defined. For
instance, when a $3\sigma$ ``glueball'' signal is discovered over an
appreciable background in a mass histogram with 100 bins, the chance
likelihood is not the nominal 0.0027 associated with a $3\sigma$
fluctuation but rather the complement of the probability that none of
the 100 bins contain such a signal, which is $1 -0.9973^{100} =
0.24$. The smaller likelihood is relevant only if we have an {\it a
priori} reason to expect that the signal would appear in the very bin
in which it was discovered.

In assessing the possibility of statistical fluctuations as the
explanation of the poor SM fit, it should be clear that the global fit
$CL$'s are appropriately defined, reflecting statistical ensembles
that correspond to replaying the previous decade of experiments many
times over. In particular, the $\chi^2$ $CL$'s of the global fits are
like the glueball example with the significance normalized to the
probability that the signal might emerge in any of the 100 bins, as
shown explicitly below.

Table 3.1 summarizes $\chi^2$ fits of four different data sets, in
which none, one, or both sets of anomalous measurements are
excluded. Consider for instance fit B in which only $x_W^{\rm
OS}[\stackrel {(-)}{\nu} N]$ is excluded, with $CL=0.10$.  In that
fit, consisting of 16 measurements, the only significantly anomalous
measurement is $A_{FB}^b$, with a pull of 2.77, for which the nominal
$CL$ is 0.0056. The likelihood that at least one of 16 measurements
will differ from the fit by $\geq 2.77\sigma$ is then $1-0.9944^{16} =
0.09$, which matches nicely with the $\chi^2$ $CL$ of 0.10. Similarly,
in fit C which retains NuTeV while excluding the hadronic asymmetry
measurements, the outstanding anomaly is $x_W^{\rm OS}[\stackrel
{(-)}{\nu} N]$ with a pull of 3.0, and the probability for at least
one such deviant is $1-0.9973^{14} = 0.04$, compared to the $\chi^2$ $CL=
0.05$. Finally, for the full data set, fit A, the outstanding anomalies
are $x_W^{\rm OS}[\stackrel {(-)}{\nu} N]$ with a pull of 3.0 and
$A_{FB}^b$ with a pull of 2.55. In that case we ask for the
probability of at least one measurement diverging by $\geq 3.0\sigma$
and a second by $\geq 2.55\sigma$, which is given by $1 -0.9973^{17} -
17(1 - 0.9973)(0.9892)^{16}= 0.006$, compared to the $\chi^2$ $CL$ of
0.010.

We see then that the $\chi^2$ $CL$'s appropriately reflect ``the
number of bins in the histogram,'' and that the poor $CL$'s of these
SM fits are well accounted for by the appropriately defined
probabilities that the outlying anomalous measurements could have
occurred by chance.  The nominal $\chi^2$ confidence levels of the
global fits are then reasonable estimates of the probability that
statistical fluctuations can explain the anomalies, which we may
characterize as unlikely but not impossible. Only fit D, with both
anomalies removed, has a robust confidence level, $CL=0.65$.  We refer
to fit D as the ``Minimal Data Set.'' The results and pulls for 
this fit are shown in table 3.2.

This discussion does not reflect the fact that the anomalous
measurements are all within the subset of measurements that dominate
the determination of $m_H$. In that smaller subset of measurements,
the significance of the anomalies is not fully reflected by the global
$CL$'s. As concerns the reliability of the fits of $m_H$, there is a
clear {\it a priori} reason to focus on the $m_H$-sensitive
measurements. We therefore also consider fits in which the observables
$O_i$ in equation (4.1) below are restricted to the measurements which
dominate the determination of $m_H$. These are the six asymmetries,
$m_W$, $\Gamma_Z$, $R_l$, and $x_W^{\rm OS}[\stackrel {(-)}{\nu}
N]$. (The $m_H$-insensitive measurements which are omitted from these
fits are $\sigma_h$, $R_b$, $R_c$, $A_b$, and $A_c$.) The results of
the corresponding fits, A$^\prime$ - D$^\prime$, are tabulated at the
bottom of table 3.1. Except for the minimal data sets, D and
D$^\prime$, in every other case the fit restricted to $m_H$-sensitive
measurements has an appreciably smaller confidence level than the
corresponding global fit. In addition to the problems of the
global fits, the poorer consistency of this sector of measurements provides
another cause for concern in assessing the reliability of the SM
prediction of $m_H$.

\noindent \underline{{\em 3.2 New Physics}}  

Each anomaly could certainly be the result of new physics. The NuTeV
experiment opens a very different window on new physics than the study
of on-shell $Z$ boson decays at LEP I and SLC.\cite{nutev_np} For
example, a $Z^\prime$ boson mixed very little or not at all with the
$Z$ boson, could have little effect on on-shell $Z$ decay but a big
effect on the NuTeV measurement, which probes a space-like region of
four-momenta centered around $Q^2 \simeq -20$ GeV$^2$. The strongest
bounds on this possibility would come from other off-shell probes,
such as atomic parity violation, $e^+e^-$ annihilation above the $Z$
pole, and high energy $p\overline p$ collisions.

New physics could also affect the hadronic asymmetry measurements.
Here we can imagine two  scenarios, depending on how
seriously we take the clustering of the three hadronic asymmetry
measurements. Taking it seriously, we would be led to consider 
leptophobic $Z^\prime$ models\cite{leptophobic}, as were invoked
to explain the $R_b$ anomaly, which was subsequently found to have a
systematic, experimental explanation.

Or we might regard the clustering of the three hadronic measurements
as accidental. Then we would be led to focus on $A_{FB}^b$, by far the
most precise of the three hadronic asymmetry measurements, and we
could arrive at acceptable global fits by assuming new physics coupled
predominantly to the third generation quarks. New physics would then
account for the $A_{FB}^b$ anomaly, with an additional effect on the
less precisely measured jet charge asymmetry, $Q_{FB}$. The third
generation is a plausible venue for new physics, since the large top
quark mass suggests a special connection of the third generation to
new physics associated with the symmetry breaking sector.

Since $R_b \propto g_{bL}^2 + g_{bR}^2$ while $A_{FB}^b\propto
g_{bL}^2 - g_{bR}^2$, and because $R_b$, which is more precisely
measured than $A_{FB}^b$, is only $\sim 1\sigma$ from its SM value,
some tuning of the shifts $\delta g_{bL}$ and $\delta g_{bR}$ is
required to fit both measurements. The right-handed coupling must
shift by a very large amount, with $\delta g_{bR} \gg \delta g_{bL}$
and $\delta g_{bR} \gtap 0.1g_{bR}$. An effect of this size suggests
new tree-level physics or radiative corrections involving a strong
interaction.  

Examples of tree-level physics are $Z\ -\ Z^\prime$ mixing or $b\ -\
Q$ mixing. A recent proposal to explain the $A_{FB}^b$ anomaly embeds
a $Z^\prime$ boson in a right-handed $SU(2)_R$ extension of the SM
gauge group in which the third generation fermions carry different
$SU(2)_R$ charges than the first and second
generations.\cite{valencia} $Z^\prime$ bosons coupled preferentially
to the third generation are generic in the context of topcolor
models.\cite{topcolor} An explanation by $b\ -\ Q$
mixing, requires $Q$ to be a charge $-1/3$ quark with non-SM weak
quantum numbers; this possibility has been explored in the context of
the latest data in reference \cite{wagner} and previously
in \cite{ma}.

If new physics explains the $A_{FB}^b$ anomaly, it must also
affect $A_b$. If we use eq. (2.3) with the factor
$A_e=0.1501(17)$ taken from the three leptonic asymmetry measurements
(assuming lepton universality), we find that the experimental value
$A_{FB}^b=0.0994(17)$ implies $A_b[A_{FB}^b]=0.883(18)$, which is
$2.89\sigma$ from $A_b[{\rm SM}]=0.935$, $CL=0.004$. However $A_b$ is
measured more directly at SLC by means of $A_{FBLR}^b$, the
front-back left-right asymmetry. In the Summer of 1998
that measurement yielded $A_b[A_{FBLR}^b]=0.867(35)$, lower by
$1.9\sigma$ than $A_b[{\rm SM}]$, lending support to the new physics
hypothesis. But the current measurement,
$A_b[A_{FBLR}^b]=0.922(20)$, is only $0.6\sigma$ below $A_b[{\rm
SM}]$. It no longer bolsters the new physics hypothesis but it is also
not grossly inconsistent with $A_b[A_{FB}^b]$, from which it differs
by $1.44\sigma$, $CL=0.15$.  Combining $A_b[A_{FB}^b]$ and
$A_b[A_{FBLR}^b]$ we find $A_b[A_{FBLR}^b \oplus
A_{FBLR}^b]=0.900(13)$, which differs from the SM by $2.69\sigma$,
$CL=0.007$. Thus while the $A_{FBLR}^b$ measurement no longer
supports the new physics hypothesis, it also does not definitively
exclude it.
 
If either anomaly is the result of new physics, the SM fails and we
cannot predict the Higgs boson mass until the nature of the new
physics is understood. New physics affecting the NuTeV
measurement and/or the hadronic asymmetry measurements will certainly
change the relationship between those observables and the value of
$m_H$, and could affect other observables in ways that
change their relationships with $m_H$.

\noindent \underline{{\em 3.3 Systematic Uncertainties}} 

The two $3\sigma$ anomalies each involve subtle systematic issues,
having to do both with performance of the measurements and the
interpretation of the results. With respect to interpretation, both
use nonperturbative QCD models with uncertainties that are difficult
to quantify. In both cases the experimental groups have put great
effort into understanding and estimating the systematic
uncertainties. Here we only summarize the main points.

The central value for the NuTeV SM result is $x_W^{\rm OS}(\stackrel
{(-)}{\nu} N)= 0.2277 \pm 0.0013({\rm statistical}) \pm 0.0009({\rm
systematic})$.  The estimated systematic error consists in equal parts of an
experimental component, $\pm 0.00063$, and a modelling component, $\pm
0.00064$. Uncertainty from the $\nu_e$ and $\overline \nu_e$ fluxes makes the
largest single contribution to the experimental component, $\pm
0.00039$, with the remainder comprised of various detector-related
uncertainties. The modelling uncertainty is dominated by
nonperturbative nucleon structure, with the biggest component, $\pm
0.00047$, due to the charm production cross section.

Two possible nonperturbative effects have been considered (see
\cite{nutev_sys,nutev_np,nutev_sys_2} and references therein). One is
an asymmetry in the nucleon strange quark sea, $\int x(s(x) -
\overline s(x))\: dx \neq 0$.  Using dimuons from the separate $\nu$
and $\overline \nu$ beams, the NuTeV collaboration finds evidence for
a $-10\%$ asymmetry within the NuTeV cross section model. If truly
present, it would increase the discrepancy from 3.0 to
3.7$\sigma$.\cite{nutev_sys} For consistency with the SM, an asymmetry
of $\simeq +30\%$ would be needed.

A second possible nonperturbative effect is isospin symmetry
breaking in the nucleon wave function, $d^p(x) \neq u^n(x)$.  Studies
are needed to determine if structure functions can be constructed that
explain the NuTeV anomaly in this way while maintaining consistency
with all other constraints. A negative result could rule out this
explanation, while a positive result would admit it as a possibility.
Confirmation would then require additional evidence.

The $3\sigma$ discrepancy between $x_W^l[A_L]$ and $x_W^l[A_H]$ also
raises the possibility of subtle systematic uncertainties.  The
determination of $x_W^l[A_L]$ from the three leptonic measurements,
$A_{LR}$, $A_{FB}^l$, and $A_{e,\tau}$, involve three quite different
techniques so that large, common systematic errors are very unlikely.
The focus instead is on the hadronic measurements, $A_{FB}^b$,
$A_{FB}^c$, and $Q_{FB}$.  In these measurements, $b$ and $\overline
c$ quarks are mutual backgrounds for one another. The signs of
both the $A_{FB}^b$ and $A_{FB}^c$ anomalies are consistent with
misidentifying $b \leftrightarrow \overline c$, although the estimated
magnitude\cite{elsing} of the effect is far smaller than what is needed. 
QCD models of charge flow and gluon radiation
are a potential source of common systematic uncertainty for all three
measurements.  The two heavy flavor asymmetries, $A_{FB}^b$ and 
$A_{FB}^c$, have the largest error correlation of the heavy flavor
$Z$-pole measurements, quoted as 16\% in the most recent
analysis.\cite{ewwg_02}

Since $x_W^l[A_H]$ is dominated by $x_W^l[A_{FB}^b]$, the greatest concern
is the systematic uncertainties of $A_{FB}^b$. The combined
result of the four LEP experiments is $A_{FB}^b= 0.0994 \pm
0.00157({\rm statistics}) \pm 0.00071({\rm systematic})$. The
systematic component arises from an ``internal'' (experimental)
component of $\pm 0.00060$ and a ``common'' component of $\pm
0.00039$, where the latter is dominated by the $\pm 0.00030$
uncertainty ascribed to QCD corrections.\cite{elsing}

\noindent \underline{{\em 3.4 Summary}}

It is not now possible to choose among the three generic
explanations of the anomalies, except to say that statistical
fluctuations are unlikely per the nominal $CL$ of the global fit.
Bigger systematic errors could rescue the SM fit but would have to be
{\em much} bigger than current estimates. Rather than further
refinement of existing error budgets, this probably means discovering
new, previously unconsidered sources of error. In this paper we focus
on the systematic error hypothesis, {\em not} because we think it the
most likely explanation --- we do not --- but, assuming it to be true, 
to see if it can really reconcile the SM with the data.

\newpage
\noindent {\bf 4. Methods}

In this section we describe the methodology of the SM 
fits. We also discuss the choice of observables, which differs 
slightly from the choice made in \cite{ewwg_02}. 

We use ZFITTER v6.30\cite{zfitter} to compute the SM electroweak radiative
corrections, with results that agree precisely
(to 2 parts in $10^5$ or better) with those obtained in
\cite{ewwg_02}. The input parameters are $m_Z$,
$m_t$, the hadronic contribution to the renormalization of $\alpha$ at
the $Z$-pole, $\Delta \alpha_5(m_Z)$, the strong coupling constant at
the $Z$-pole, $\alpha_S(m_Z)$, and the Higgs boson mass, $m_H$.
For any point in this five dimensional space ZFITTER provides the
corresponding SM values of the other observables, $O_i$, listed in
table 2.1. 

To generate the $\chi^2$ distributions we scan over the four parameters,  
$m_t$, $\Delta \alpha_5(m_Z)$, $\alpha_S(m_Z)$, and $m_H$. For a specified  
collection of observables $O_i$, we then have 
$$
\chi^2= \left({m_t - m_t^{\rm EXPT} \over \Delta m_t^{\rm EXPT}}\right)^2
       + \left({\Delta \alpha_5- \Delta \alpha_5^{\rm EXPT}
                 \over \Delta(\Delta \alpha_5^{\rm EXPT})}\right)^2
       + \sum_i\left({O_i-O_i^{\rm EXPT} \over \Delta O_i^{\rm EXPT}}\right)^2
                                        \eqno{(4.1)}
$$
The experimental values are given in table 2.1.

We do not scan over $m_Z$ but simply fix it at its central
experimental value.  Because $m_Z$ is so much more precise than the
other observables, it would contribute negligibly to $\chi^2$
if we did scan on it. We have verified this directly by performing
fits in which it was varied, with only negligible differences from the
fits in which it is fixed at the central value.  This can also be seen
in the global fits reported by the EWWG, in which the pull from $m_Z$
is invariably much less than 1. In this case, inclusion of $m_Z$ has
no effect on the $CL$ of the fit, because (1) the contribution to
$\chi^2$ is negligible and (2) the scan on $m_Z$ has no effect on the
number of degrees of freedom since it is both varied and
constrained. Since it has little effect, we choose not to scan on
$m_Z$ in order to facilitate the numerical calculations.

For $\Delta \alpha_5(m_Z)$ we use the determination of \cite{bp}, 
which incorporates the most recent $e^+e^-$ annihilation data
and is also the default choice of \cite{ewwg_02}. In \cite{mc1} we
also presented results for four other determinations of $\Delta
\alpha_5(m_Z)$.

For the global fits $\alpha_S(m_Z)$ is left unconstrained, as is also
done in \cite{ewwg_02}, because the $Z$-pole SM fit is itself the most
precise determination of $\alpha_S(m_Z)$. For the fits which consider
more limited sets of observables, we use the following rule: if at
least two of the three observables which dominate the the
determination of $\alpha_S(m_Z)$ (these are $\Gamma_Z$, $R_l$, and
$\sigma_h$) are included in the fit, $\alpha_S$ is unconstrained as in
the global fits. Otherwise we constrain it to 0.118(3). In any case,
because the $m_H$-sensitive observables are predominantly
$\alpha_S$-insensitive, the results we obtain for $m_H$ depend very
little on the details of how $\alpha_S$ is specified.

The fits also include the important correlations from the error
matrices presented in tables 2.3 and 5.3 of \cite{ewwg_02}. We retain
the correlations that are $\geq 0.05$ in the $6\times 6$ heavy flavor
error matrix, for $A_{FB}^b$, $A_{FB}^c$, $A_b$, $A_c$, $R_b$, and
$R_c$. Similarly we retain correlations $\geq 0.05$ in the $4 \times
4$ correlation matrix for $\Gamma_Z$, $\sigma_h$, $R_l$, and
$A_{FB}^l$. These correlations shift the value of $\chi^2$ by as
much as 2 units.

Our global fits differ slightly from the all-data fit of
\cite{ewwg_02}, principally because we use the set of measurements
the Electroweak Working Group (EWWG) used through
Spring 2001 but not two measurements that were subsequently
added by the EWWG.  From Summer 2001 the EWWG all-data fit
included the Cesium atomic parity violation (APV) measurement, and
in Spring 2002, the $W$ boson width, $\Gamma_W$, was included. 

Reference \cite{ewwg_02} uses a 2001 determination of the Cesium APV
measurement that has recently been superseded by newer results from
the same authors\cite{dfg_02}. With the average value from the more
recent study, $Q_W= -72.18(46)$ (with experimental and theoretical
errors combined in quadrature), and the SM value from \cite{ewwg_02},
the pull is 1.6, rather than 0.6 as quoted in \cite{ewwg_02}.  The
effect on the all-data global fit in \cite{ewwg_02} is
to change $\chi^2/N=28.8/15$, $CL=0.017$ to $\chi^2/N=30.8/15$,
$CL=0.009$, decreasing the confidence level by a factor two.  Rather
than use the updated value, we choose not to include the Cs APV
measurement since the theoretical systematic uncertainties are still
in flux.

We choose not to include $\Gamma_W$ because it has not yet attained a
level of precision precision approaching that of the other
measurements in the fit. For instance, $\Gamma_Z$ is 30 times more
precise than $\Gamma_W$, so that $\Gamma_W$ has 1/900'th the weight of
$\Gamma_Z$ in the global fits.  At the current precision it has no
sensitivity to new physics signals of the order of magnitude probed by
the other observables in the fit.  Its effect on the fits of $m_H$,
which are the principal concern of this work, is completely
negligible.

In any case, the decision not to include the $\Gamma_W$ and APV
measurements does not have a major effect on our results.  In
particular, the effect on the Higgs boson mass predictions is
negligible. Furthermore, our all-data fit, with $CL=0.010$, has a
very similar confidence level to that of the all-data fit of
\cite{ewwg_02}, $CL=0.009$, if the APV determination is updated to
reflect \cite{dfg_02}.

We closely reproduce the results of \cite{ewwg_02} when we use the
same set of observables. For instance, adding $\Gamma_W$ and $Q_W({\rm
Cs})$ as specified in \cite{ewwg_02} to the observables in our global
fit, table 2.1, we obtain $\chi^2/N=28.7/15$, $CL= 0.018$, compared to
$\chi^2/N=28.8/15$, $CL= 0.017$ from the corresponding (all-data) fit
of \cite{ewwg_02}.

\noindent {\bf 5. The $\chi^2$ Fits}

In this section we present several SM fits of the data, using $\chi^2$
to estimate global $CL$'s and $\Delta \chi^2$ to obtain the 
constraints on the Higgs boson mass.  The $\Delta \chi^2$ method, used
also in \cite{ewwg_02}, is defined as follows. Let $m_{\rm MIN}$ be
the value of $m_H$ at the $\chi^2$ minimum, and let $m_L$ be an
arbitrary mass such that $m_L>m_{\rm MIN}$. Then the confidence level
$CL(m_H > m_L)$ is one half of the confidence level corresponding to 
a $\chi^2$ distribution for one degree of freedom, with the 
value of the $\chi^2$ distribution given by $\Delta \chi^2=
\chi^2(m_L) -\chi^2(m_{\rm MIN})$. We consider both global fits and fits
restricted to the $m_H$-sensitive observables.

In addition to $m_t$ and $\Delta \alpha_5$, which are input parameters
to the ZFITTER calculations, the observables with the greatest
sensitivity to $m_H$ are $x_W^l$ from the six asymmetry measurements,
$m_W$, $\Gamma_Z$, $R_l$, and $x_W^{\rm OS}[\stackrel {(-)}{\nu}
N]$. It is useful to consider the domains in $m_H$ favored by these
observables, in order to understand the ``alliances'' (see section 2)
that shape the global fit, and to understand the consistency of the
fits with the LEP II lower limit on $m_H$. In fits restricted to
$m_H$-sensitive obervables, $\chi^2$ is given by eq. 4.1, where the
$O_i$ are restricted to the $m_H$-sensitive observables under
consideration. In addition, for fits containing fewer than two of the
three $\alpha_S$-sensitive observables --- $\Gamma_Z$, $R_l$, and
$\sigma_h$ --- we constrain $\alpha_S(m_Z)$ by including it with the $O_i$
in equation (4.1) as discussed in section 4.

The experimental quantity that currently has the greatest sensitivity
to $m_H$ is $x_W^l$, determined from the six asymmetry
measurements. Figure 1 displays the distributions of the three
individual leptonic measurements and the combined result from all
three, $x_W^l[A_L]$; it shows that the upper limit is dominated by
$A_{LR}$.  The central value, $m_H=55$ GeV, symmetric 90\% confidence
interval, and likelihood $CL(m_H>114)$ are given in table 5.1.
Note that the 95\% upper limit is $m_H<143$ GeV.

Figure 2 shows that the $\chi^2$ distribution from $x_W^l[A_H]$ is
completely dominated by the $b$ quark asymmetry, $A_{FB}^b$.  The
central value is $m_H=410$ GeV and the 95\% lower limit is 145 GeV,
just above the 95\% upper limit from $x_W^l[A_L]$. The 95\% upper
limit extends above 1 TeV.  Figure 3 shows the $\chi^2$ distributions
of both $x_W^l[A_L]$ and $x_W^l[A_H]$, with the respective symmetric
90\% confidence intervals indicated by the dot-dashed horizontal
lines.

It is also interesting to isolate the effect of the $W$ boson mass
measurement, because it is the second most important quantity for
fixing $m_H$, and because it is a dramatically different measurement
with a completely different set of systematic uncertainties.  Figure
4 shows the $\chi^2$ distribution from $m_W$.  The central value is
$m_H=23$ GeV and the 95\% upper limit is 121 GeV. Both $m_W$ and
$x_W^l[A_L]$ then favor very light values of $m_H$. This is the basis of
the ``alliance'' between $x_W^l[A_L]$ and $m_W$, discussed in section 2,
that pushes $A_{FB}^b$ to outlyer status and 
contributes to the marginal confidence level of the SM fit.

The two other $Z$-pole observables with sensitivity to $m_H$ are
$\Gamma_Z$ and $R_l$, which also involve different systematic
uncertainties than the asymmetry measurements. The corresponding
$\chi^2$ distributions are also plotted in figure 4, together with the
combined distribution for $m_W$, $\Gamma_Z$ and $R_l$. $R_l$ also
favors small $m_H$, with its $\chi^2$ minimum off the chart below 10
GeV.  $\Gamma_Z$ is often represented as favoring $m_H \simeq {\rm
O}(100)$ GeV, but we see in figure 4 and with the expanded scale in
figure 5, that it actually has two nearly degenerate minima, at about
16 and 130 GeV.\footnote{I wish to thank D. Bardin and G. Passarino,
who kindly verified this surprising feature, using, respectively,
recent versions of ZFITTER\cite{zfitter} and TOPAZ0.\cite{topaz}} In
table 5.1 we see that the combined distribution of the three
non-asymmetry measurements has a central value at 13 GeV and a 95\%
upper limit at 73 GeV with $CL(m_H)>114)=0.021$

Figure 6 shows the $\chi^2$ distribution from the NuTeV
measurement, $x_W^{\rm OS}[\stackrel {(-)}{\nu} N]$.  The minimum lies
above 3 TeV and the 95\% lower limit is at $\simeq 660$ GeV.  The SM
fits the NuTeV anomaly by driving $m_H$ to very large
values, but the new physics that actually explains the effect, if it
is genuine, would not be so simply tied to the symmetry breaking
sector but might for instance reflect an extension of the gauge
sector, with implications for the symmetry breaking sector that cannot
be foretold. In any case, as discussed in section 7, values of 
$m_H$ above $\sim 1$ TeV cannot be interpreted literally. 

It is striking that the measurements favoring $m_H$ in the region
allowed by the direct searches are precisely the ones responsible 
for the large $\chi^2$ of the global fit. They favor values far above
100 - 200 GeV while the measurements consistent with the fit favor
values far below, as shown in the bottom two lines of table 5.1.
The fit based on $x_W^l[A_L] \oplus m_W\oplus \Gamma_Z \oplus R_l$ 
(fit $D^\prime$ in table 3.1) has $m_H<106$ GeV at 95\% $CL$,
while the fit based on $x_W^l[A_H] \oplus x_W^{\rm OS}[\stackrel
{(-)}{\nu} N]$ has $m_H>220$ GeV at 95\% $CL$.  The corresponding
$\chi^2$ distributions are shown in figure 7.
 
Next we consider the global fits that were discussed in section 3.1.
The principal results are summarized in table 5.2.  The ``all-data''
fit, fit A (shown in detail in table 2.1), closely resembles the
all-data fit of \cite{ewwg_02}, up to small differences arising from
the slightly different choice of observables discussed in section
4. As summarized in table 5.2, the $\chi^2$ minimum is at $m_H=94$
GeV, with $CL(\chi^2)= 0.010$. The 95\% upper limit is $m_H < 193$ GeV
and the consistency with the search limit is $CL(m_H>114\ {\rm
GeV})= 0.30$. The $\chi^2$ distribution is shown in figure 8, where
the vertical dashed line denotes the direct search limit and the
horizontal dot-dashed line indicates the symmetric 90\% $CL$ interval.
The combined likelihood for internal consistency of the fit and
consistency between fit and search limit, defined in equation (1.1),
is $P_C= 0.0030$.

The ``Minimal Data Set,'' fit D, with $x_W^{\rm OS}[\stackrel
{(-)}{\nu} N]$ and $x_W^l[A_H]$ omitted, is shown in detail in table
3.2, and the $\chi^2$ distribution is shown in figure 8.  The minimum
is at $m_H=43$ GeV, with a robust confidence level $CL(\chi^2)=
0.65$. But the 95\% upper limit is $m_H < 105$ GeV and the consistency
with the search limit is a poor $CL(m_H>114\ {\rm GeV})= 0.035$.
(The latter is nearly identical to the value 0.038 shown in table 5.1
for fit D$^\prime$, which is the corresponding fit restricted to
$m_H$-sensitive observables, defined in section 3.1.) The combined
likelihood for fit D is $P_C= 0.023$.

In fit C with $x_W^l[A_H]$ omitted the $\chi^2$ minimum is at $m_H=45$
GeV with $CL(\chi^2)= 0.066$. The confidence level for consistency
with the search limit is $CL(m_H>114\ {\rm GeV})= 0.047$. The combined
likelihood is $P_C= 0.0031$.

Finally we consider fit B, with $x_W^l[A_H]$ retained and $x_W^{\rm
OS}[\stackrel {(-)}{\nu} N]$ omitted. Now the $m_H$ prediction is
raised appreciably, with the $\chi^2$ minimum at $m_H=81$ GeV, but the
quality of the fit is marginal, with $CL(\chi^2)= 0.10$. The
confidence level for consistency with the search limit is more robust,
$CL(m_H>114\ {\rm GeV})= 0.26$, and the combined likelihood is
$P_C= 0.026$.

The effect of the hadronic asymmetries and the NuTeV measurement is
apparent from table 5.2. The NuTeV measurement diminishes $CL(\chi^2)$
by a factor 10, seen by comparing fit A with fit B and fit C with fit
D, while its effect on $CL(m_H>114)$ is modest. Consequently the NuTeV
measurement also diminishes the combined likelihood $P_C$ by an
order of magnitude. Comparing fit A with fit C or fit B with fit D, we
see that the hadronic asymmetries also diminish $CL(\chi^2)$, by a
factor $\sim 7$, but that they increase $CL(m_H>114)$ by a nearly
identical factor, so that they have little effect on $P_C$.

\noindent {\bf 6. ``Bayesian'' Maximum Likelihood Fits}

The $\Delta \chi^2$ method for obtaining the confidence levels for
different regions of $m_H$ is poweful and convenient but, at least to
this author, not completely transparent. We have therefore also
approached these estimates by constructing likelihood
distributions as a function of $m_H$,
varying the parameters to find the point of maximum likelihood for each
value of $m_H$. Assuming Gaussian statistics, the log likelihood is
$$
-{\rm log}({\cal L}(m_H)) = C \chi^2(m_H),    \eqno{(6.1)}
$$
so that the maximization of the likelihood is equivalent to the
minimization of $\chi^2$. 
The proportionality constant $C$ is determined by the normalization
condition for ${\cal L}$.  

The method is ``Bayesian'' in the sense that the domain of
normalization and the measure are specified by {\it a priori} choices
that are guided by the physics. The likelihood distribution is
normalized in the interval $10\: {\rm GeV} \: \leq \: m_H\:
\leq\: 3 {\rm TeV}$, 
$$
\int^{m_H=3\: {\rm TeV}}_{m_H=10 {\rm GeV}} d{\rm log}(m_H) {\cal L}(m_H) 
              = 1.            \eqno{(6.2)}   
$$ 
The choice of measure is motivated by the fact that ${\rm log}(m_H)$
is approximately linearly proportional to the experimental parameters,
which are assumed to be Gaussian distributed, such as $m_t$, $\Delta
\alpha_5$ and the various $O_i$ --- see for instance the interpolating
formulae in \cite{dgps}.  The choice of interval is conservative in
the sense that enlarging the domain above and below causes $CL(m_H>114
{\rm GeV})$ to be even smaller than the results given below.

The normalized likelihood distributions for fits A and D, the all-data
and Minimal Data Sets, are shown in figure 9, where we display both
the differential and integrated distributions. The confidence level
$CL(m_H>114)$ is the area under the differential distribution
above 114 GeV.  For the Minimal Data Set the result is
$CL(m_H>114\ {\rm GeV})= 0.030$, in good agreement with the result
0.035 obtained from $\Delta \chi^2$ in section 5.  For the all-data
set it is $CL(m_H>114\ {\rm GeV})= 0.25$, compared with 0.30 from
$\Delta \chi^2$.

It is clear from figure 9 that the likelihood distribution from the
Minimal Data Set is vanishingly small at 3 TeV but has some support at
10 GeV. If we were to enlarge the domain in $m_H$ both above and
below, the effect would be to further decrease the likelihood for 
$m_H>114$ GeV.

\noindent {\bf 7. New Physics in the Oblique Approximation}

If we assume the Minimal Data Set, the contradiction with the LEP
II lower limit on $m_H$ is either a statistical fluctuation or a
signal of new physics. Two recent papers provide examples of new
physics that could do the job. Work by Altarelli {\it et
al.}\cite{altarellietal} in the framework of the MSSM finds that the
prediction for $m_H$ can be raised into the region allowed by the
Higgs boson searches if there are light sneutrinos, $\simeq 55$ -- 80
GeV, light sleptons, $\simeq 100$ GeV, and moderately large tan$\:
\beta \simeq 10$.  This places the sleptons just beyond the
present experimental lower limit, where they could be discovered in
Run II at the TeVatron.  A second proposal, by Novikov {\it et
al.},\cite{novikovetal} finds that a fourth generation of quarks and
leptons might also do the job, provided the neutrino has a mass just
above $m_Z/2$. An illustrative set of parameters is $m_N \simeq 50$
GeV, $m_E \simeq 100$ GeV, $m_U + m_D \simeq 500$ GeV, $|m_U - m_D|
\simeq 75$ GeV, and $m_H \simeq 300$ GeV.

In this section we do not focus on any specific model of new physics
but consider the class of models that can be represented in the
oblique approximation\cite{oblique}, parameterized by the 
quantities $S, T, U$.\cite{pt} The essential assumption is that the
dominant effect of the new physics on the electroweak observables can
be parameterized as effective contributions to the $W$ and $Z$ boson
self energies. These contributions are not limited to loop
corrections, since the oblique parameters can also 
represent tree level phenomena such as $Z - Z^\prime$
mixing.\cite{holdom} We will restrict ourselves to the $S$ and $T$
parameters, since they suffice to make the point that oblique new
physics can remove the contradiction between the Minimal Data Set and
the search limit, leaving $m_H$ as an essentially free
parameter. We also show that $S,T$ corrections do not 
improve the confidence levels of the global fits that
include the anomalous measurements.

For the observables $O_i$ the oblique corrections are given by
$$
dQ_i= \sum_i (A_i S + B_i T)       \eqno(7.1)
$$
with $Q_i$ defined by $Q_i = O_i$ or $Q_i = {\rm ln} (O_i)$ as
indicated in table 7.1 where $A_i$ and $B_i$ are tabulated. Since
these are not SM fits, instead of $x_W^{\rm OS}[\stackrel {(-)}{\nu}
N]$ the NuTeV experiment is represented by the model independent fit
to the effective couplings $g_L^2= g_{uL}^2 + g_{dl}^2$ and $g_R^2=
g_{uR}^2 + g_{dR}^2$, for which the experimental values from
\cite{nutev} are given in table 2.1.

Figure 10 shows the $S,T$ fit to the Minimal Data Set along with the 
SM fit with $S=T=0$. The striking feature of the $S,T$ fit is that 
$\chi^2$ is nearly flat as a function of $m_H$. There is therefore 
no problem reconciling the fit with the lower limit on $m_H$, and 
there is also no preference for any value of $m_H$. The fits are 
acceptable all the way to $m_H=3$ TeV, and the variation across the 
entire region is bounded by $\Delta \chi^2 \leq 1.2$. Because the 
minimum is so shallow, it is not significant that it occurs 
at $m_H=17$ GeV. The confidence level 
at the minimum is 0.51, which is comparable to the confidence level, 
$CL=0.65$, of fit D, the corresponding SM fit.

It is well known that arbitrarily large values of $m_H$ can be
accomodated in $S,T$ fits of the electroweak data.\cite{flat} This can
be understood as a consequence of the fact the SM fit of $m_H$ is
dominated by two observables, $x_W^l$ and $m_W$.  Let $m_H$[MIN],
$x_W^l$[MIN] and $m_W$[MIN] be the values of $m_H$, $x_W^l$ and $m_W$
at the $\chi^2$ minimum of the SM fit.  The shifts $\delta x_W^l$ and
$\delta m_W$ induced in the SM fit by choosing a different value of
$m_H \neq m_H$[MIN], can then be compensated by choosing $S$ and $T$
to provide equal and opposite shifts, $-\delta x_W^l$ and $-\delta
m_W$. Inverting the expressions from equation (3.13) of \cite{pt} we
have explicitly, in the approximation that we consider only $x_W^l$
and $m_W$,
$$
S= -{4\over \alpha}\left(\delta x_W^l +2x_W^l {\delta m_W \over m_W}
          \right) 		\eqno(7.2)
$$
and 
$$
T= -{2\over \alpha (1 - x_W^l)}\left(\delta x_W^l +
          {\delta m_W \over m_W}\right).	\eqno(7.3)
$$
For instance, for $m_H=1$ TeV the $S,T$ fit to the Minimal Data Set
(set D) shown in figure 10 yields $S,T= -0.22,+0.34$ compared with
$S,T= -0.15,+0.22$ from equations (7.2) and (7.3). The approximation
correctly captures the trend though it differs by 30\% from the
results of the complete fit. The discrepancy reflects the importance of
variations among parameters other than $x_W^l$ and $m_W$ that are
neglected in deriving (7.2) and (7.3).

Values of $m_H$ above 1 TeV cannot be interpreted literally as
applying to a simple Higgs scalar. For $m_H>1$ TeV symmetry breaking
is dynamical, occurring by new strong interactions that cannot be
analyzed perturbatively.\cite{mcmkg} If the Higgs mechanism is
correct, there are new quanta that form symmetry breaking vacuum
condensates.  Values of $m_H$ above 1 TeV should be regarded only as a
rough guide to the order of magnitude of the masses of the
condensate-forming quanta.

It is sometimes said that an SM Higgs scalar above $\simeq 600$ GeV
is excluded by the triviality bound, which is of order 1 TeV in
leading, one loop order\cite{dn}, refined to $\simeq 600$ GeV in
lattice simulations.\cite{triv_lattice} The bound is based on
requiring that the Landau singularity in the Higgs boson
self-coupling, $\lambda$, occur at a scale $\Lambda_{\rm Landau}$ that
is at least twice the Higgs boson mass, $\Lambda_{\rm Landau} \gtap
2m_H$, in order for the SM to have some minimal ``head room'' as an
effective low energy theory.  However, the conventional analysis does
not include the effect on the running of $\lambda$ from the new
physics which {\em must} exist at the Landau singularity.  Although
power suppressed, in the strong coupling regime, which is in fact the
relevant one for the upper limit on $m_H$, the power suppressed
corrections can change the predicted upper limit appreciably, possibly
by factors of order one.\cite{mc_triv} To take literally the 600 GeV
upper bound from lattice simulations we in effect assume that the new
physics is a space-time lattice. The bound cannot be known precisely
without knowing something about the actual physics that replaces the
singularity.  The analysis in \cite{mc_triv} is performed in the
symmetric vacuum and should be reconsidered for the spontaneously
broken case, but the conclusion is likely to be unchanged since it
follows chiefly from the ultraviolet behavior of the effective theory
which is insensitive to the phase of the vacuum. An SM scalar between
600 GeV and 1 TeV therefore remains a possibility.

Figure 10 also displays the values of $S$ and $T$ corresponding 
to the $\chi^2$ minimum at each value of $m_H$. For $m_H > 114$ GeV 
the minima fall at moderately positive $T$ and moderately negative $S$. 
Positive $T$ occurs naturally in models that break custodial $SU(2)$, 
for instance with nondegenerate quark or lepton isospin doublets. 
Negative $S$ is less readily obtained but there is not a no-go 
theorem, and models of new physics with $S<0$ have been 
exhibited.\cite{negative_s}

We also consider a fit to the Minimal Data Set in which only 
$T$ is varied with $S$ held at $S=0$. The result is shown in figure 11. The
minimum falls at $m_H=55$ GeV with $CL= 0.56$. The distribution at
larger $m_H$ is flat though not as flat as the $S,T$ fit.  Moderately
large, postive $T$ is again preferred. From $\Delta \chi^2$ we find
that the confidence level for $m_H$ above the LEP II lower limit is
sizeable, $CL(m_H>114\: {\rm GeV})= 0.21$, and that the 95\% upper
limit is $m_H < 460$ GeV.

Next we consider the $S,T$ fit with the hadronic asymmetry
measurements included, corresponding to SM fit B above. Shown
in figure 12, it is not improved relative to the SM fit. The $\chi^2$
minimum is at $m_H=15$ GeV with $\chi^2/N= 15/10$ implying
$CL=0.12$. From $\Delta \chi^2$ the probability for $m_H$ in the
allowed region is a marginal $CL(m_H>114\: {\rm GeV})= 0.08$, and
the combined probability from equation (1.1) is $P_C= 0.01$.

The all-data $S,T$ fit, including both the hadronic asymmetry and
NuTeV measurements, is shown in figure 13. In this case the $S=T=0$
fit is not identical to SM fit A, since the NuTeV result is
parameterized by $g_L$ and $g_R$ rather than $x_W^{\rm OS}[\stackrel
{(-)}{\nu} N]$ as in the SM fit. The minimum of the $S=T=0$ fit occurs
at $m_H=94$ GeV, with $\chi^2/N= 26/14$ implying $CL=0.026$. The $S,T$
fit is actually of poorer quality: the shallow minimum is at $m_H= 29$
GeV with $\chi^2/N= 25.7/12$ implying $CL=0.012$.

\newpage
\noindent {\bf 8. Discussion}

Taken together the precision electroweak data and the direct searches
for the Higgs boson create a complex puzzle with many possible
outcomes. An overview is given in the ``electroweak schematic
diagram,'' figure 14. The diagram illustrates how various hypotheses
about the two $3\sigma$ anomalies lead to new physics or to the
conventional SM fit. The principal conclusion of this paper is
reflected in the fact that the only lines leading into the `SM' box 
are labeled `Statistical Fluctuation.' That is, systematic error
alone cannot save the SM fit, since it implies the conflict with the
search limit, indicated by the box labeled $CL(m_H>114)=0.035$, which
in turn either implies new physics or itself reflects statistical
fluctuation. This is a consequence of the fact that the combined
probability $P_C$ defined in equation (1.1) is poor whether the
anomalous measurements are included in the fit or not, as summarized
in table 5.2.

The `New Physics' box in figure 14 is reached if either $3\sigma$
anomaly is genuine or, conversely, if neither is genuine and the
resulting 96.5\% $CL$ conflict with the search limit is
genuine.  It is also possible to invoke statistical fluctuation as the
exit line from any of the three central boxes. However we have argued
that the global confidence levels indicated for fits A and B are fair
reflections of the probability that those fits are fluctuations from
the Standard Model. As such they do not favor the SM while they also
do not exclude it definitively: ``It is a part of probability that
many improbable things will happen.''\cite{aristotle}

The smoothest path to the SM might be the one which traverses the
central box, fit B, and then exits via `Statistical Fluctuation' to
the SM. In this scenario nucleon structure effects might explain the
NuTeV anomaly and the 10\% confidence level of fit B could be a
fluctuation. This is a valid possibility, but two other problems
indicated in the central box should also be considered in evaluating
this scenario. First, the consistency of the $m_H$-sensitive
measurements is even more marginal, indicated by the 3.4\% confidence
level of fit B$^\prime$. Second, the troubling $3\sigma$
conflict ($CL=0.003$) between the leptonic and hadronic asymmetry
measurements is at the heart of the determination of $m_H$. Thus
even if we assume that the marginal $CL$ of the global fit is due to
statistical fluctuation, the reliability of the prediction of $m_H$
hangs on even less probable fluctuations. As noted
above, to be consistent with the search limit statistical
fluctuations must involve both the `anomalous' hadronic asymmetry
measurements {\em and} the measurements that conform to the SM fit,
especially the leptonic asymmetry measurements and the $W$ boson mass
measurement.  The conflict with the search limit would be greatly
exacerbated if the true value of $x_W^l[A_H]$ were equal to the
present value of $x_W^l[A_L]$.

Since there are still some ongoing analyses of the hadronic asymmetry
data, there may yet be changes in the final results, but unless major
new systematic effects are uncovered the changes are not likely to be
large. More precise measurements might be made eventually at a second
generation Z factory, such as the proposed Giga-Z project.  However,
to fully exploit the potential of such a facility it will be necessary
to improve the present precision of $\Delta \alpha_5(m_Z)$ by a factor
of $\sim 5$ or better, requiring a dedicated program to improve our
knowledge of $\sigma(e^+e^- \rightarrow {\rm HADRONS})$ below $\sim 5$
GeV.\cite{jegerlehner} The $W$ boson and top quark mass measurements
will be improved at Run II of the TeVatron, at the LHC, and eventually
at a linear $e^+e^-$ collider. For instance, an upward shift of the
top quark mass\cite{mt_effect} or a downward shift of the $W$ boson
mass could diminish the inconsistency between the Minimal Data Set and
the search limit, while shifts in the opposite directions would
increase the conflict.\footnote{The probability of such shifts is
of course encoded in the fits by the contributions of $m_W$ and $m_t$ to
$\chi^2$.}

The issues raised by the current data set heighten the excitement of
this moment in high energy physics.  The end of the decade of
precision electroweak measurements leaves us with a great puzzle, that
puts into question the mass scale at which the physics of electroweak
symmetry breaking will be found. The solution of the puzzle could
emerge in Run II at the TeVatron.  If it is not found there it is very
likely to emerge at the LHC, which at its design luminosity will be
able to search for the new quanta of the symmetry breaking sector over
the full range allowed by unitarity.

\vskip 0.2in

\noindent {\bf Acknowledgements:} I am grateful to Martin Grunewald for
his kind and prompt responses to my questions about the EWWG SM
fits. I also thank Kevin McFarland and Geralyn Zeller for useful
correspondence and comments. I thank Dimitri Bardin and Giampiero
Passarino for kindly verifying the peculiar dependence of $m_H$ on
$\Gamma_Z$, Robert Cahn for classical references, and Max Chanowitz
for preparing figure 14.


\newpage
\vskip 0.5in

\noindent Table 2.1. SM All-data fit (fit A). Experimental values for the
model-independent parameters $g_L^2[\stackrel {(-)}{\nu} N]$ and
$g_R^2[\stackrel {(-)}{\nu} N]$ are given for completeness but are not
used in the SM fit. 

\begin{center}
\vskip 12pt
\begin{tabular}{c|ccc}
 &Experiment& SM Fit& Pull \\ 
\hline
\hline
$A_{LR}$ & 0.1513 (21)  & 0.1481  &1.6  \\
$A_{FB}^l$ & 0.0171 (10) &0.0165  & 0.7 \\
$A_{e,\tau}$ & 0.1465 (33) & 0.1481 & -0.5 \\
$A_{FB}^b$ & 0.0994 (17) & 0.1038 &-2.6  \\
$A_{FB}^c$ & 0.0707 (34) & 0.0742 & -1.0 \\
$x_W^l[Q_{FB}]$ & 0.2324 (12) & 0.23139 & 0.8  \\
$m_W$ & 80.451 (33) & 80.395 & 1.7 \\
$\Gamma_Z$ & 2495.2 (23) & 2496.4 &-0.5  \\
$R_l$ & 20.767 (25) &20.742  &1.0  \\
$\sigma_h$ & 41.540 (37) & 41.479 &1.6  \\
$R_b$ & 0.21646 (65) & 0.21575 &1.1  \\
$R_c$ & 0.1719 (31) & 0.1723 &-0.1  \\
$A_b$ & 0.922 (20) & 0.9347 &-0.6  \\
$A_c$ & 0.670 (26) &  0.6683 & 0.1 \\
$x_W^{\rm OS}[\stackrel {(-)}{\nu} N]$ & 0.2277 (16) &0.2227  &3.0  \\
$m_t$ & 174.3 (5.1) &175.3  &-0.2  \\
$\Delta \alpha_5(m_Z^2)$ & 0.02761 (36) &0.02768  & 0.2 \\
$\alpha_S(m_Z)$ &  &0.1186  &  \\
$m_H$ & & 94 &\\
\hline
$g_L^2[\stackrel {(-)}{\nu} N]$ &  0.3005 (14)&  &  \\
$g_R^2[\stackrel {(-)}{\nu} N]$ & 0.0310 (11) &  &  \\
\hline
\hline
\end{tabular}
\end{center}

\newpage
\vskip 0.5in

\noindent Table 2.2. Evolution of the electroweak data. As noted in the text, 
the same data is tracked for the three data sets though, following 
\cite{ewwg_02}, it is grouped into fewer degrees of freedom in the Spring 
'02 data set. 
\begin{center}
\vskip 12pt
\begin{tabular}{c|ccc}
& Summer '98& Spring '01& Spring '02 \\
\hline
\hline
$x_W^l[A_L]$ &0.23128 (22) &0.23114 (20) &0.23113 (21) \\
$x_W^l[A_h]$ &0.23222 (33) &0.23240 (29) &0.23220 (29) \\
$CL(A_L \oplus A_H)$ &0.02 &0.0003 &0.003 \\
$CL(x_W^l)$ &0.25 &0.02 &0.06 \\
$m_W$  &80.410 (90) &80.448 (34) &80.451 (33) \\
$\chi^2/N$ (no $x_W^{\rm OS}[\stackrel {(-)}{\nu} N]$&13.8/14&24.6/14&18.4/12\\
$CL[\chi^2/N]$  &0.46 &0.04 &0.10 \\
\hline
$x_W^{\rm OS}[\stackrel {(-)}{\nu} N]$&0.2254 (21) &0.2255 (21) &0.2277 (16) \\
Pull($x_W^{\rm OS}[\stackrel {(-)}{\nu} N]$)  &1.1 &1.2 &3.0 \\
$\chi^2/N$  &15/15 &26/15 &27.7/13 \\
$CL[\chi^2/N]$  &0.45 &0.04 &0.01 \\
\hline
\hline
\end{tabular}
\end{center}

\vskip 0.5in

\noindent Table 2.3. ``What if?'': role of $m_W$ in shaping the global
fit.  The first column reflects actual current data with $x_W^{\rm
OS}[\stackrel {(-)}{\nu} N]$ omitted. In the second and third columns 
$m_W$ is assigned hypothetical values as described in the text, while other 
measurements are held at their Spring '02 values. 
In each case the effect of omitting $A_{FB}^b$ or $A_{LR}$ is also 
shown.

\begin{center}
\vskip 12pt
\begin{tabular}{c|ccc}
& $m_W['02]$&  $m_W['98]$&  $\Delta m_W['02]$\\
\hline
\hline
$m_W$ & 80.451 (33) & 80.410 (90) & 80.370 (33) \\
\hline
$\chi^2/12$,  $CL$ &18.4, 0.10 &15.2, 0.23 & 15.3, 0.23 \\
\hline
$- A_{FB}^b$ &&&\\
$\chi^2/11$, $CL$ & 10.2, 0.51 & 9.0, 0.62 & 9.8, 0.55 \\
{\bf OR} &&&\\
 $-A_{LR}$ &&&\\
$\chi^2/11$, $CL$ & 15.7, 0.15 & 10.2, 0.51 & 10.0, 0.53 \\
\hline
\hline
\end{tabular}
\end{center}

\vskip 1in

\noindent Table 3.1. Results for global fits A - D and for the
corresponding fits restricted to $m_H$-sensitive observables, 
A$^\prime$ - D$^\prime$.

\begin{center}
\vskip 12pt
\begin{tabular}{c||c|c}
 & All &$- x_W^{\rm OS}[\stackrel {(-)}{\nu} N]$ \\
\hline
\hline
  All & \bf A & \bf B \\
    & $\chi^2/=27.7/13$, $CL=0.010$ & 18.4/12, 0.10 \\
\hline
  $-x_W^l[A_H]$ & {\bf C} & \bf D \\
                & 17.4/10, 0.066  &  6.8/9, 0.65 \\
\hline
\hline
$m_H$-sensitive only: & &  \\
All & \bf A$^\prime$ & \bf B$^\prime$ \\
    & 24.3/8, 0.0020 &  15.2/7, 0.034 \\
\hline 
  $-x_W^l[A_H]$ & {\bf C$^\prime$} & \bf D$^\prime$ \\
    & 13.8/5, 0.017 & 3.45/4, 0.49 \\
\hline
\hline
\end{tabular}
\end{center}

\vskip .5in

\noindent Table 3.2. SM fit D, to Minimal Data Set, with 
$x_W^{\rm OS}[\stackrel {(-)}{\nu} N]$ and three hadronic asymmetry 
measurements excluded. 

\begin{center}
\vskip 12pt
\begin{tabular}{c|ccc}
 &Experiment& SM Fit& Pull \\ 
\hline
\hline
$A_{LR}$ & 0.1513 (21)  & 0.1509  & 0.2  \\
$A_{FB}^l$ & 0.0171 (10) &0.0171  & 0.0 \\
$A_{e,\tau}$ & 0.1465 (33) & 0.1509 & -1.4 \\
$m_W$ & 80.451 (33) & 80.429 & 0.7 \\
$\Gamma_Z$ & 2495.2 (23) & 2496.1 &-0.4  \\
$R_l$ & 20.767 (25) &20.737  &1.2  \\
$\sigma_h$ & 41.540 (37) & 41.487 & 1.4  \\
$R_b$ & 0.21646 (65) & 0.21575 &1.1  \\
$R_c$ & 0.1719 (31) & 0.1722 &-0.1  \\
$A_b$ & 0.922 (20) & 0.9350 &-0.7  \\
$A_c$ & 0.670 (26) &  0.670 & 0.0 \\
$m_t$ & 174.3 (5.1) &175.3  &-0.2  \\
$\Delta \alpha_5(m_Z^2)$ & 0.02761 (36) &0.02761  & 0.0 \\
$\alpha_S(m_Z)$ &  &0.1168  &  \\
$m_H$ & & 43 &\\
\hline
\hline
\end{tabular}
\end{center}

\vskip .5in

\noindent Table 5.1. Predictions for $m_H$ from various restricted
sets of $m_H$-sensitive observables. The value of $m_H$ at the
$\chi^2$ minimum is shown along with the symmetric 90\% confidence
interval and the likelihood for $m_H>114$ GeV.  Values indicated as
$10-$ or $3000+$ fall below or above the interval $10 <m_H<3000$ GeV
within which the fits are performed.

\begin{center}
\vskip 12pt
\begin{tabular}{c|ccc}
 & $m_H$ (GeV) & 90\% $CL$ & $CL(m_H>114)$ \\ 
\hline
\hline
$x_W^l[A_L]$ & 55 & $16 <m_H<143$ & 0.10 \\
$x_W^l[A_H]$ & 410 & $145 <m_H< 1230$ & 0.98 \\
$m_W$ & 23 & $10- <m_H< 121$ & 0.059\\
$m_W\oplus \Gamma_Z \oplus R_l$ & 13 &$ 10- <m_H< 73$ & 0.021 \\
$x_W^{\rm OS}[\stackrel {(-)}{\nu} N]$ &$3000+$ & $ 660 < m_H<3000+$&0.996 \\
\hline
$x_W^l[A_L] \oplus m_W\oplus \Gamma_Z \oplus R_l$ & 37 & $ 11 <m_H<106 $
                       & 0.038\\
$x_W^l[A_H] \oplus x_W^{\rm OS}[\stackrel {(-)}{\nu} N]$&600& $220<m_H<1690 $
                                   & 0.995\\
\hline
\hline
\end{tabular}
\end{center}

\vskip .5in

\noindent Table 5.2. Confidence levels and Higgs boson mass
predictions for global fits A - D. Each entry shows the value of $m_H$
at the $\chi^2$ minimum, the symmetric 90\% confidence interval, the
$\chi^2$ confidence level, the confidence level for consistency with
the search limit, and the combined likelihood $P_C$ from
equation (1.1).

\begin{center}
\vskip 12pt
\begin{tabular}{c||c|c}
 & All &$- x_W^{\rm OS}[\stackrel {(-)}{\nu} N]$ \\
\hline
\hline
  All & \bf A & \bf B \\
   & $m_H= 94$ & $m_H= 81$  \\  
   & $37<m_H<193$ & $36<m_H<190$  \\              
   &$CL(\chi^2)=0.010$ & $CL(\chi^2)=0.10$  \\
   & $CL(m_H>114)=0.30$ & $CL(m_H>114)=0.26$  \\
   & $P_C= 0.0030$ & $P_C= 0.02$6 \\
\hline
  $-x_W^l[A_H]$ & {\bf C} & \bf D \\
   & $m_H= 45$ & $m_H= 43$  \\              
   & $14<m_H<113$ & $17<m_H<105$  \\              
   & $CL(\chi^2)=0.066$ &  $CL(\chi^2)=0.65$  \\
   & $CL(m_H>114)=0.047$ &  $CL(m_H>114)=0.035$  \\
   & $P_C= 0.0031$ & $P_C= 0.023$ \\
\hline
\hline
\end{tabular}
\end{center}

\vskip 1in

\noindent Table 7.1 Coefficients for the oblique corrections as defined 
in equation (7.1).

\begin{center}
\vskip 12pt
\begin{tabular}{c|cc}
  $Q_i$ & $A_i$ & $B_i$ \\
\hline
\hline
 $A_{LR}$  &  -0.0284 & 0.0202 \\
 $A_{FB}^l$ & -0.00639 & 0.00454 \\
 $A_{e,\tau}$  &  -0.0284 & 0.0202 \\
 $x_W^l[Q_{FB}]$  & 0.00361    &  -0.00256   \\
 $A_{FB}^c$  & -0.0156    & 0.0111    \\
 $A_{FB}^b$  & -0.0202    & 0.0143    \\
 ln($\Gamma_Z$)  & -0.00379    &  0.0105   \\
 ln($R_l$)   & -0.00299    & 0.00213    \\
 ln($\sigma_h$)  & 0.000254    & -0.000182    \\
 $m_W$  & -0.00361    & 0.00555    \\
 ln($R_c$)  & -0.00127    &  0.000906   \\
 ln($R_b$)  & 0.000659    &  -0.000468   \\
 $A_c$  & -0.0125    &  0.00886   \\
 $A_b$  &   -0.00229  & 0.00163    \\
 $g_L^2$  & -0.00268    &  0.00654   \\
 $g_R^2$  & 0.000926    & -0.000198    \\
\hline
\hline
\end{tabular}
\end{center}

\newpage
\begin{figure}

\begin{center}                                                                 
\includegraphics[height=6in,width=4in,angle=90]{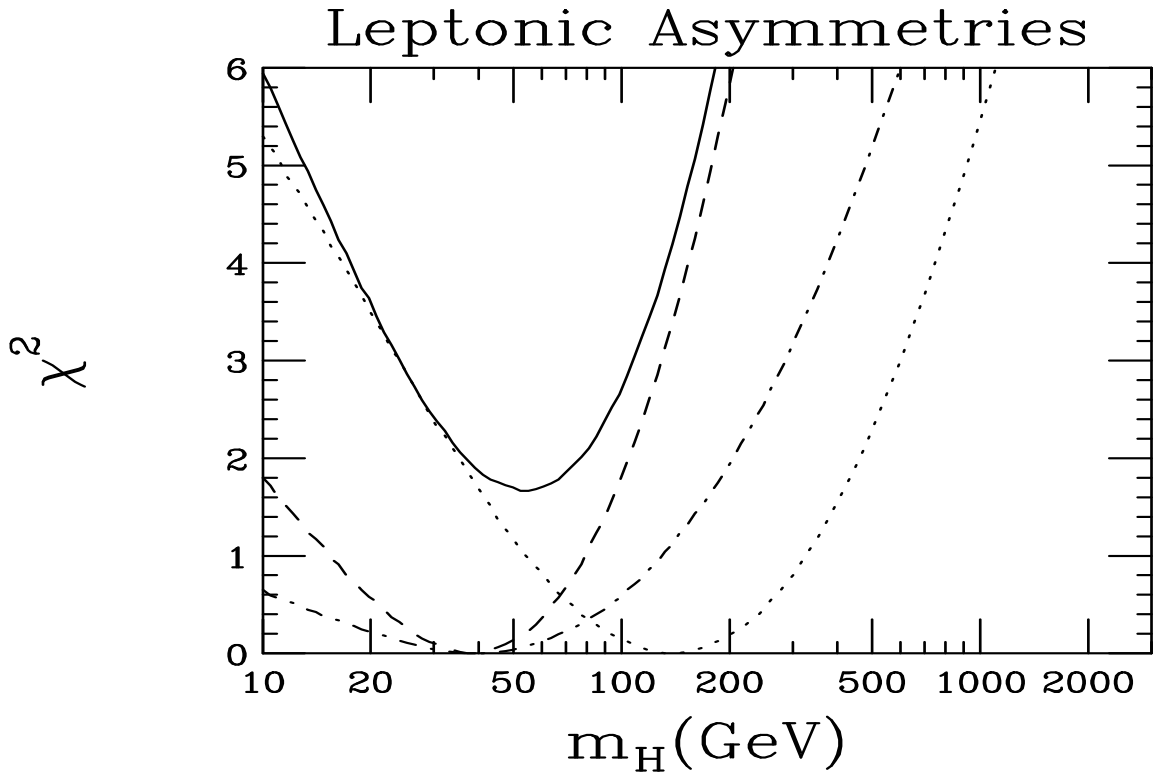}
\end{center}

\caption{$\chi^2$ distributions as a function of $m_H$ from the
leptonic asymmetry measurements. The dashed, dot-dashed, and dotted
lines are obtained from $A_{LR}$, $A_{FB}^l$, and $A_{e,\tau}$
respectively. The solid line is the combined fit to the three
asymmetries.}

\label{fig1}
\end{figure}

\newpage
\begin{figure}

\begin{center}                                                                 
\includegraphics[height=6in,width=4in,angle=90]{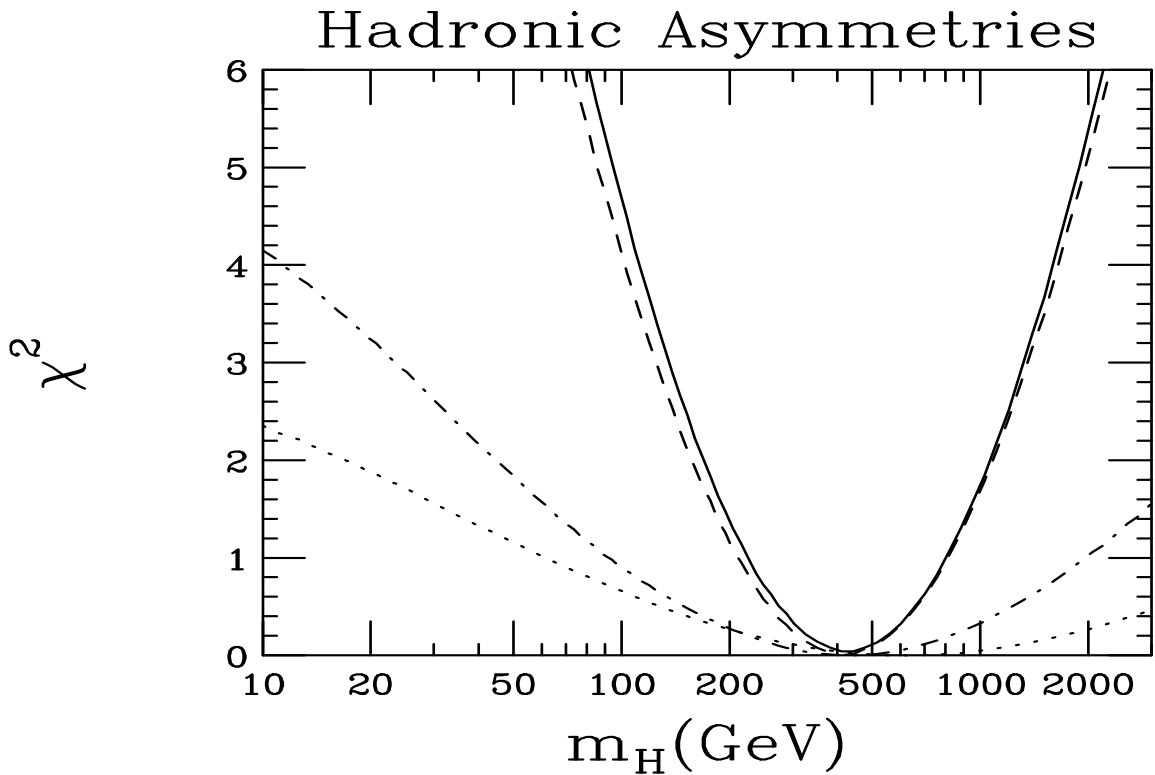}
\end{center}

\caption{$\chi^2$ distributions from the hadronic asymmetry 
measurements. The dashed, dot-dashed, and dotted lines are 
obtained from $A_{FB}^b$, $A_{FB}^c$, and $Q_{FB}$ respectively. The 
solid line is the combined fit to the three asymmetries.}

\label{fig2}
\end{figure}

\newpage
\begin{figure}

\begin{center}                                                                 
\includegraphics[height=6in,width=4in,angle=90]{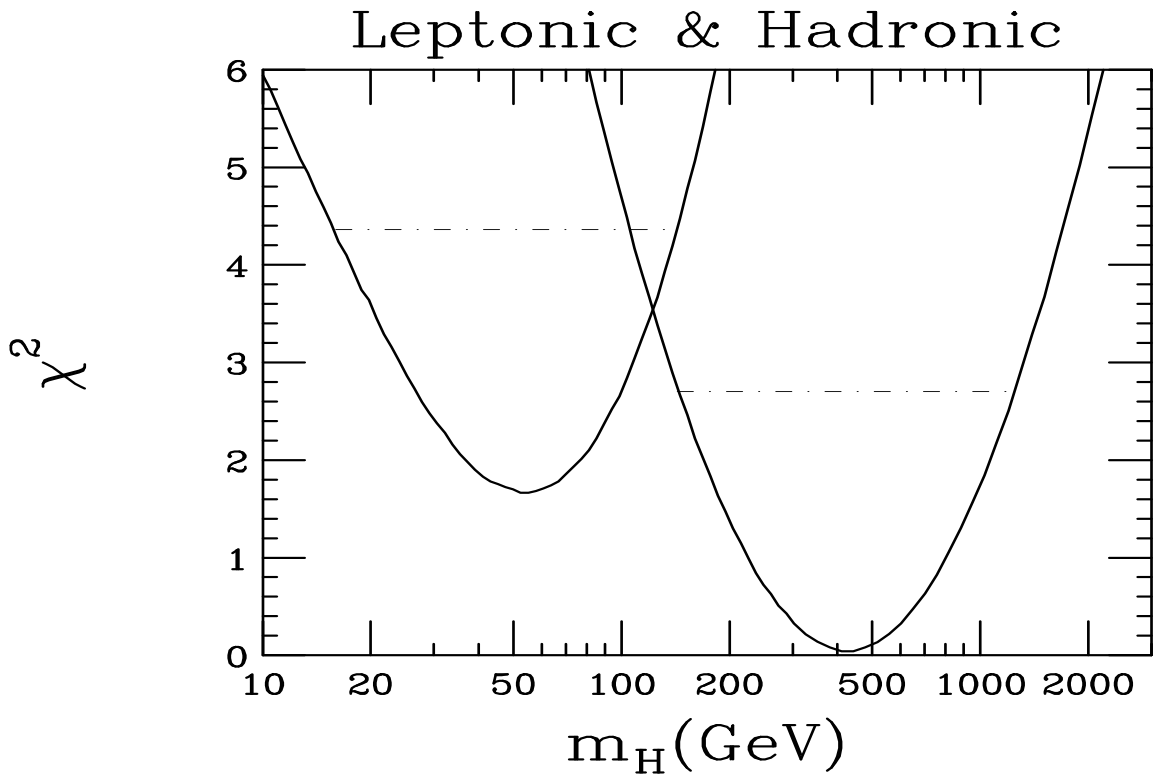}
\end{center}

\caption{$\chi^2$ distributions from the leptonic (left) and hadronic
(right) asymmetry measurements. The dot-dashed lines indicate the
respective symmetric 90\% $CL$ intervals.}

\label{fig3}
\end{figure}

\newpage

\begin{figure}

\begin{center}                                                                 
\includegraphics[height=6in,width=4in,angle=90]{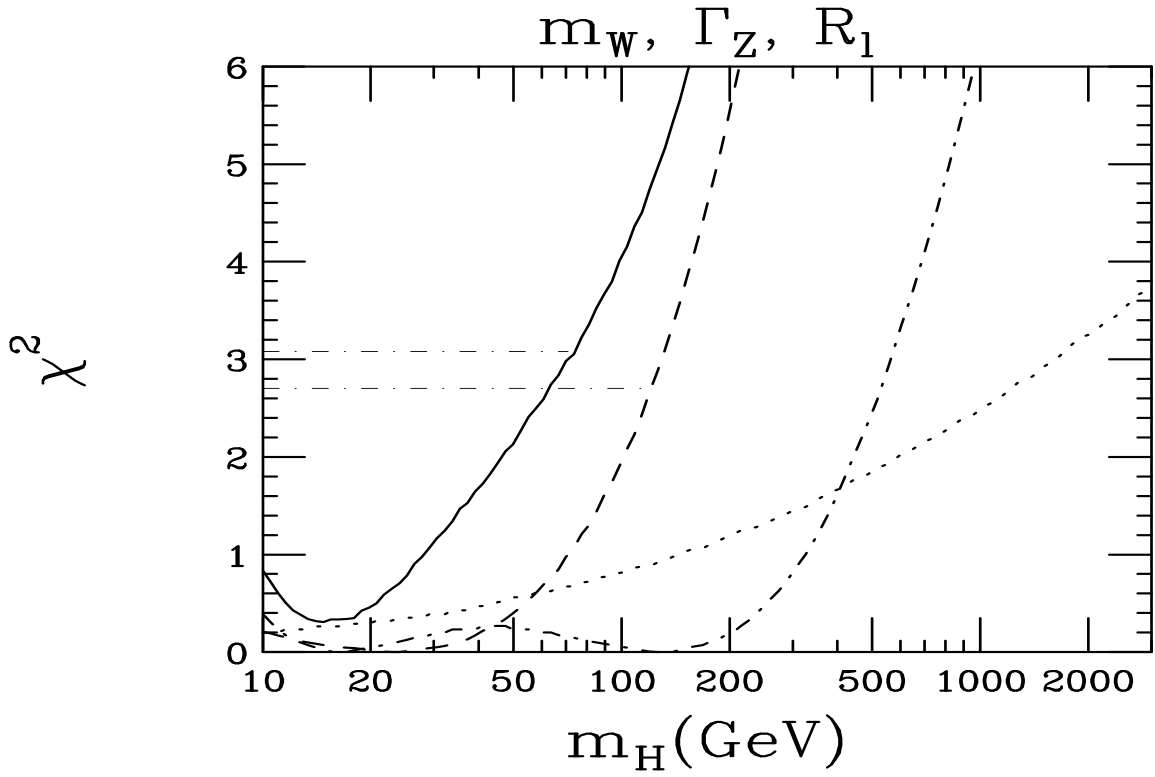}
\end{center}

\caption{$\chi^2$ distributions from non-asymmetry measurements. The
dashed, dot-dashed, and dotted lines are obtained from $m_W$,
$\Gamma_Z$, and $R_l$ respectively. The solid line is the combined fit
to the three measurementss.  The dot-dashed lines mark the 
95\% $CL$ upper limits for the combined distribution and for the 
distribution obtained from $m_W$ alone.}

\label{fig4}
\end{figure}

\newpage
\begin{figure}

\begin{center}                                                                 
\includegraphics[height=6in,width=4in,angle=90]{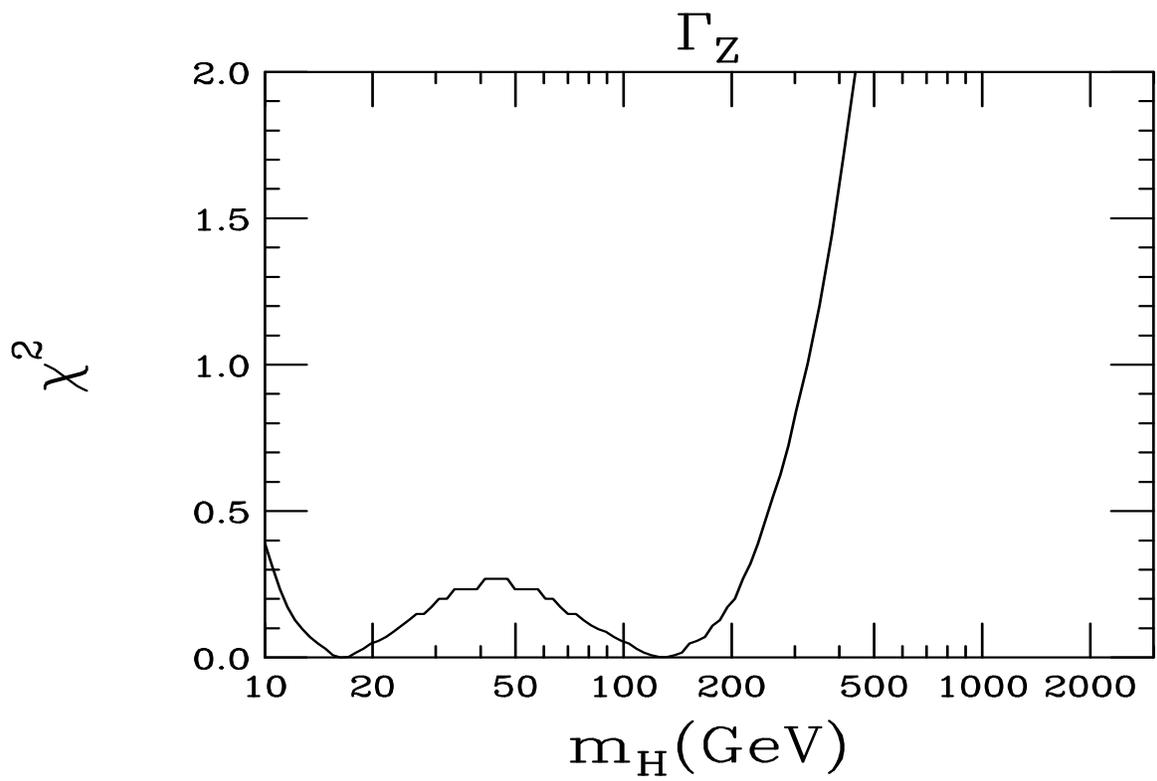}
\end{center}

\caption{$\chi^2$ distribution from $\Gamma_Z$ with expanded scale.}

\label{fig5}
\end{figure}

\newpage
\begin{figure}

\begin{center}                                                                 
\includegraphics[height=6in,width=4in,angle=90]{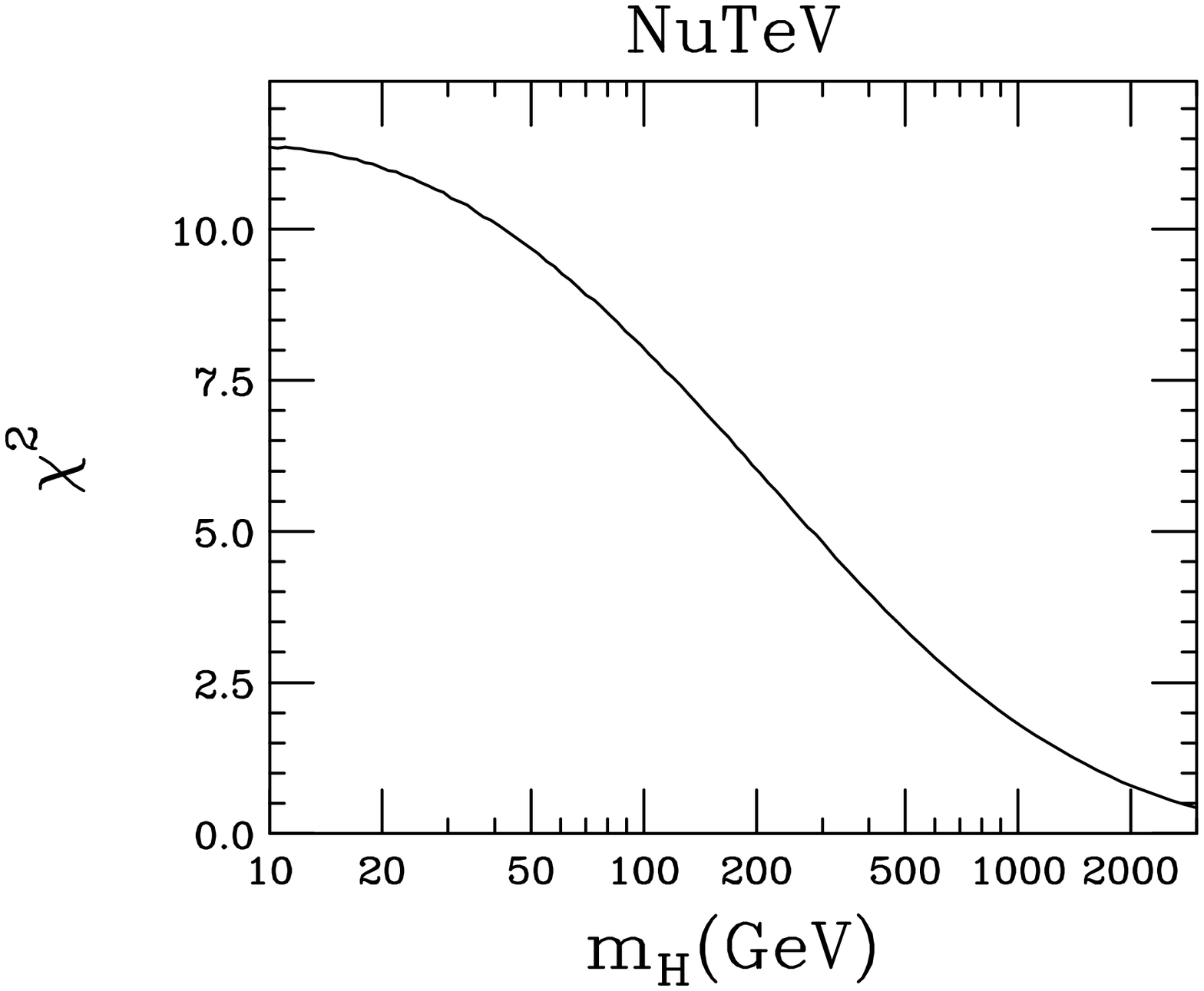}
\end{center}

\caption{$\chi^2$ distribution from $x_W^{\rm OS}[\stackrel {(-)}{\nu} N]$.}

\label{fig6}
\end{figure}

\newpage
\begin{figure}

\begin{center}                                                                 
\includegraphics[height=6in,width=4in,angle=90]{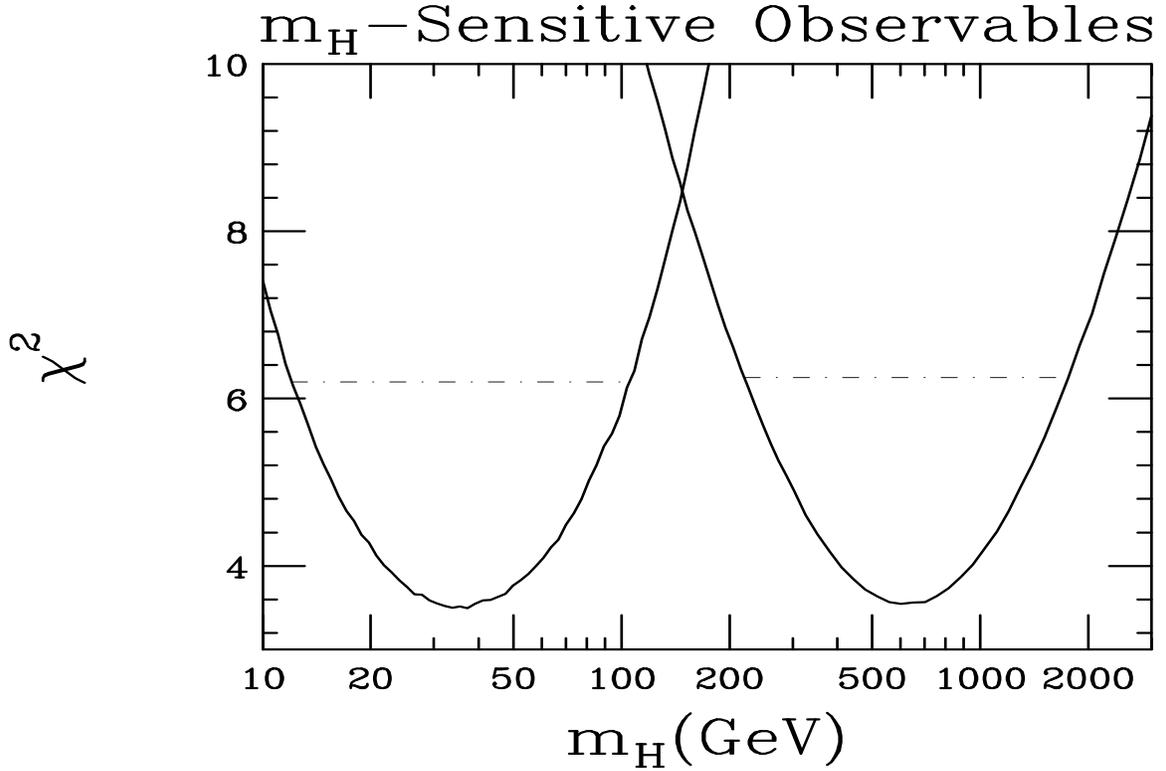}
\end{center}

\caption{$\chi^2$ distributions from the $m_H$-sensitive observables.
The distribution on the left is a fit to the D$^\prime$ data set, i.e., 
restricted to the $m_H$-sensitive observables $A_{LR}$, $A_{FB}^l$,
$A_{e,\tau}$, $m_W$, $\Gamma_Z$, and $R_l$. The distribution on the
right is a fit to the remaining $m_H$-sensitive observables:
$A_{FB}^b$, $A_{FB}^c$, $Q_{FB}$, and $x_W^{\rm OS}[\stackrel
{(-)}{\nu} N]$. The dot-dashed lines indicate the respective symmetric
90\% $CL$ intervals.}

\label{fig7}
\end{figure}

\newpage
\begin{figure}

\begin{center}                                                                 
\includegraphics[height=6in,width=4in,angle=90]{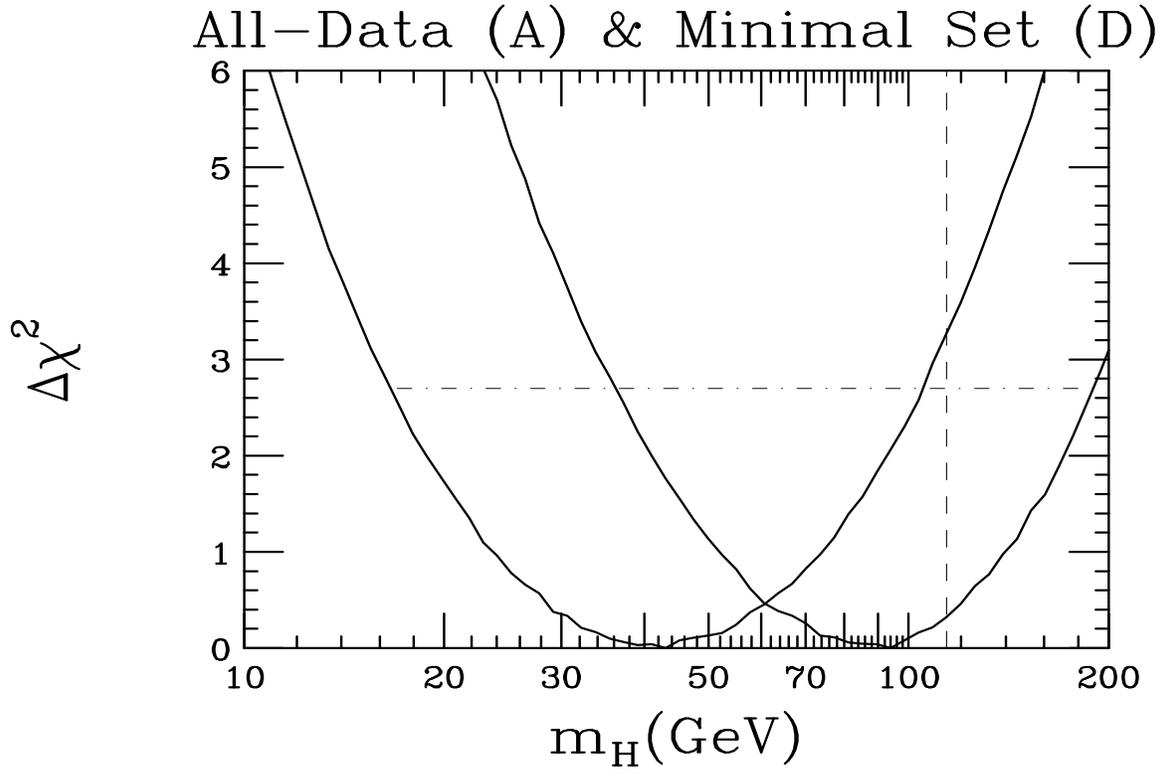}
\end{center}

\caption{$\Delta \chi^2$ for two SM global fits. The Minimal Data Set, 
fit D, is on the left and the all-data set, fit A, is on the right. The 
90\% symmetric confidence intervals are indicated by the horizontal 
dot-dashed line. The vertical dashed line denotes the 95\% $CL$ lower 
limit on $m_H$ from the direct searches.}

\label{fig8}
\end{figure}

\newpage
\begin{figure}

\begin{center}                                                                 
\includegraphics[height=6in,width=4in,angle=90]{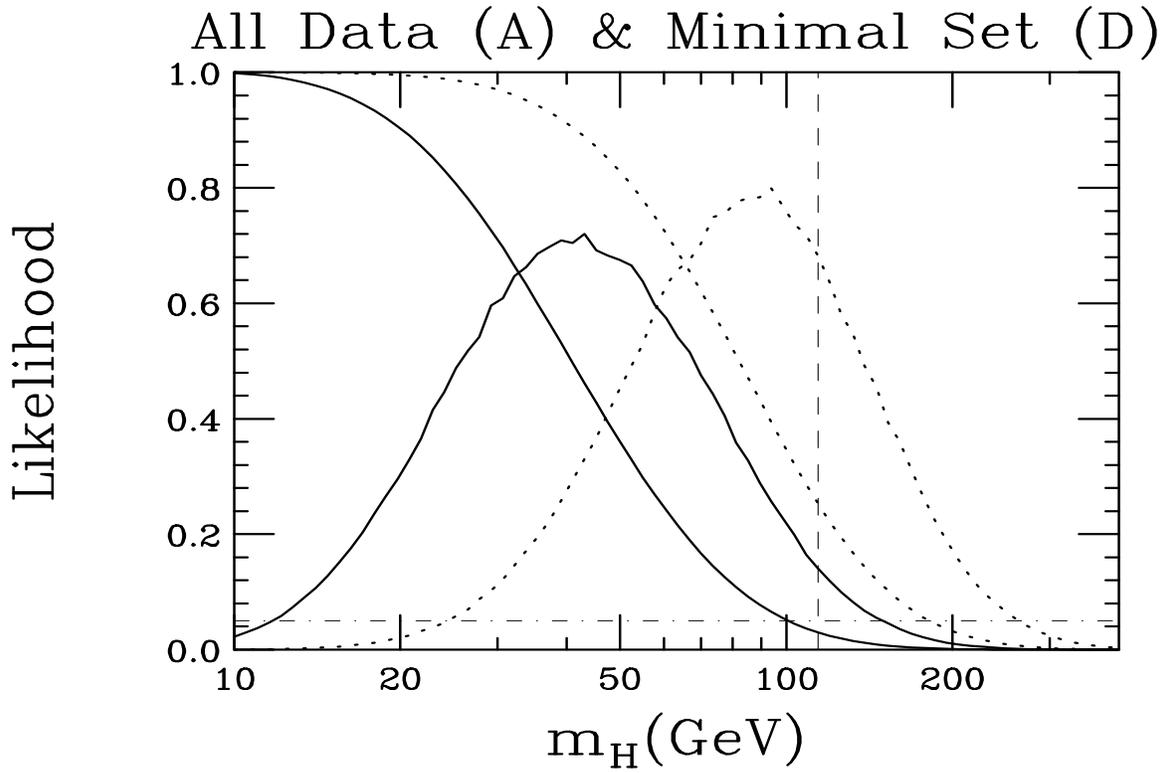}
\end{center}

\caption{Differential and integrated likelihood distributions for the 
Minimal Data Set (set D, solid lines) and the all-data set (set A, 
dotted lines). The vertical dashed line denotes the 95\% $CL$ lower 
limit on $m_H$ from the direct searches, and the horizontal dot-dashed 
line indicates the 5\% likelihood level.}

\label{fig9}
\end{figure}

\newpage
\begin{figure}

\begin{center}                                                                 
\includegraphics[height=6in,width=4in,angle=90]{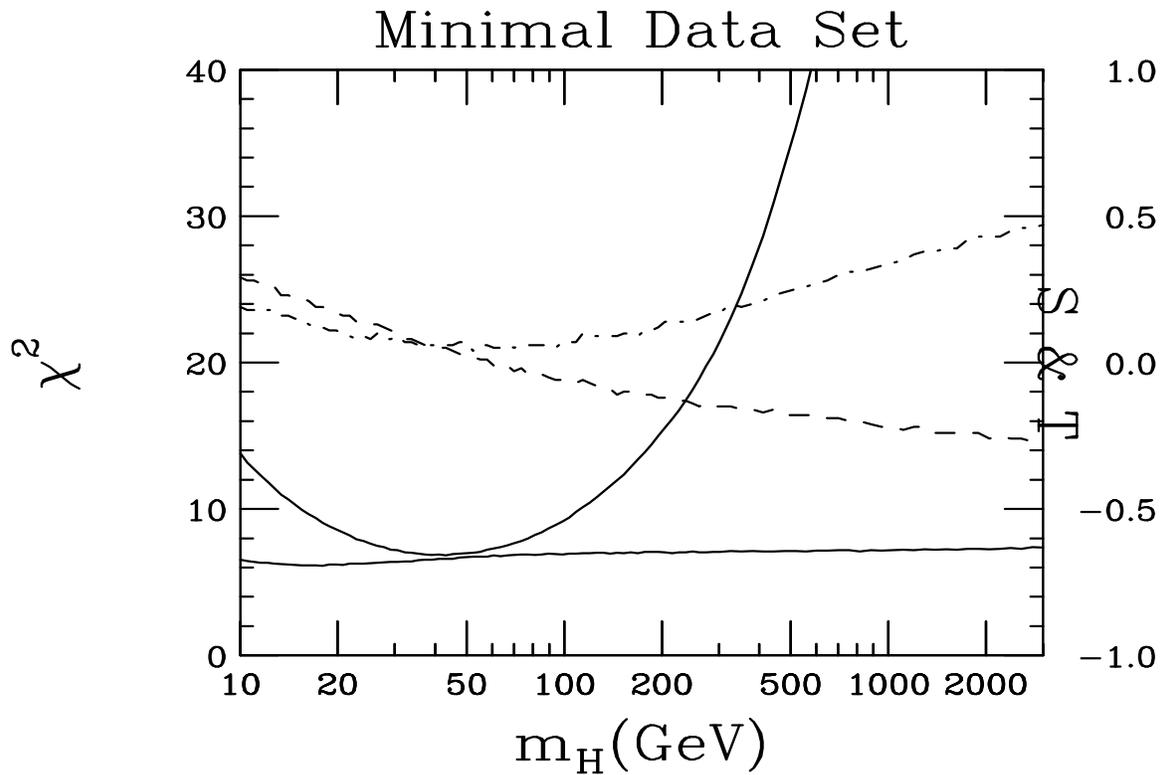}
\end{center}

\caption{$\chi^2$ distributions (solid lines) for the SM  and $S,T$ fits 
to the Minimal Data Set (data set D). The corresponding values of $S$ 
(dashed line) and $T$ (dot-dashed line) are read to the scale on the 
right axis.}

\label{fig10}
\end{figure}

\newpage
\begin{figure}

\begin{center}                                                                 
\includegraphics[height=6in,width=4in,angle=90]{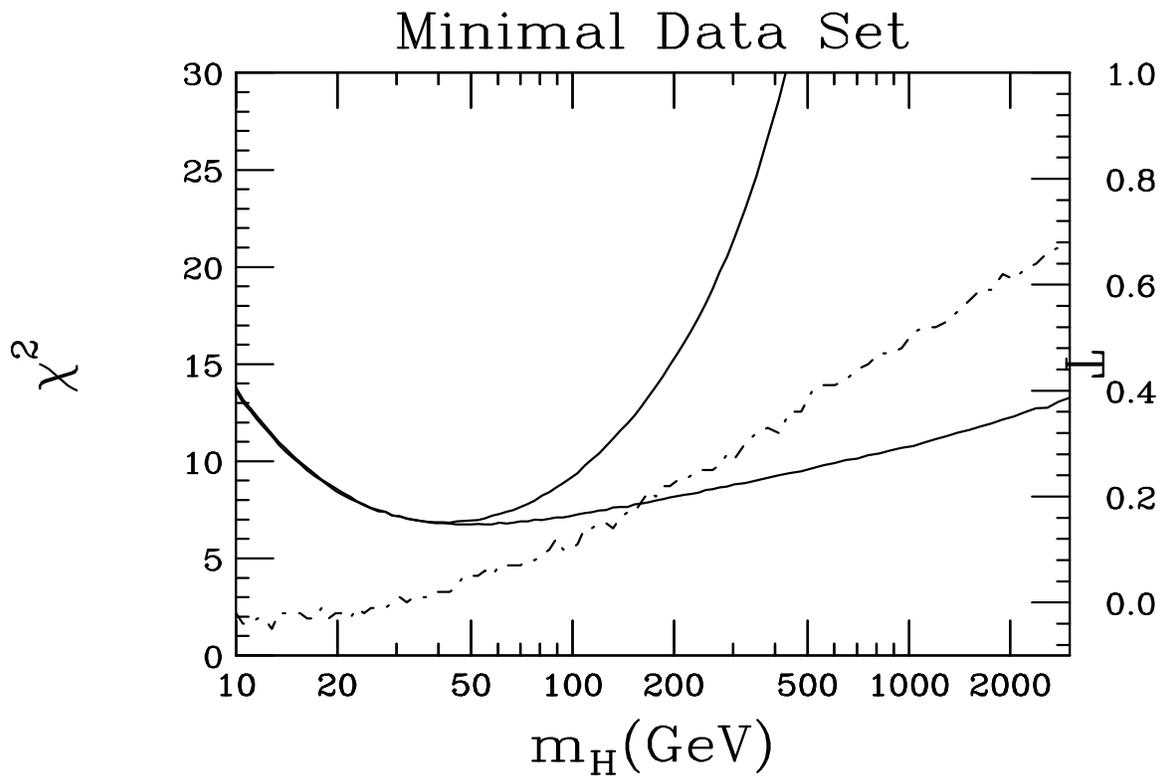}
\end{center}

\caption{$\chi^2$ distributions (solid lines) for the SM and $T$-only
fits to the Minimal Data Set (data set D). 
$T$ (dot-dashed line) is read to the scale on the right axis.}

\label{fig11}
\end{figure}

\newpage
\begin{figure}

\begin{center}                                                                 
\includegraphics[height=6in,width=4in,angle=90]{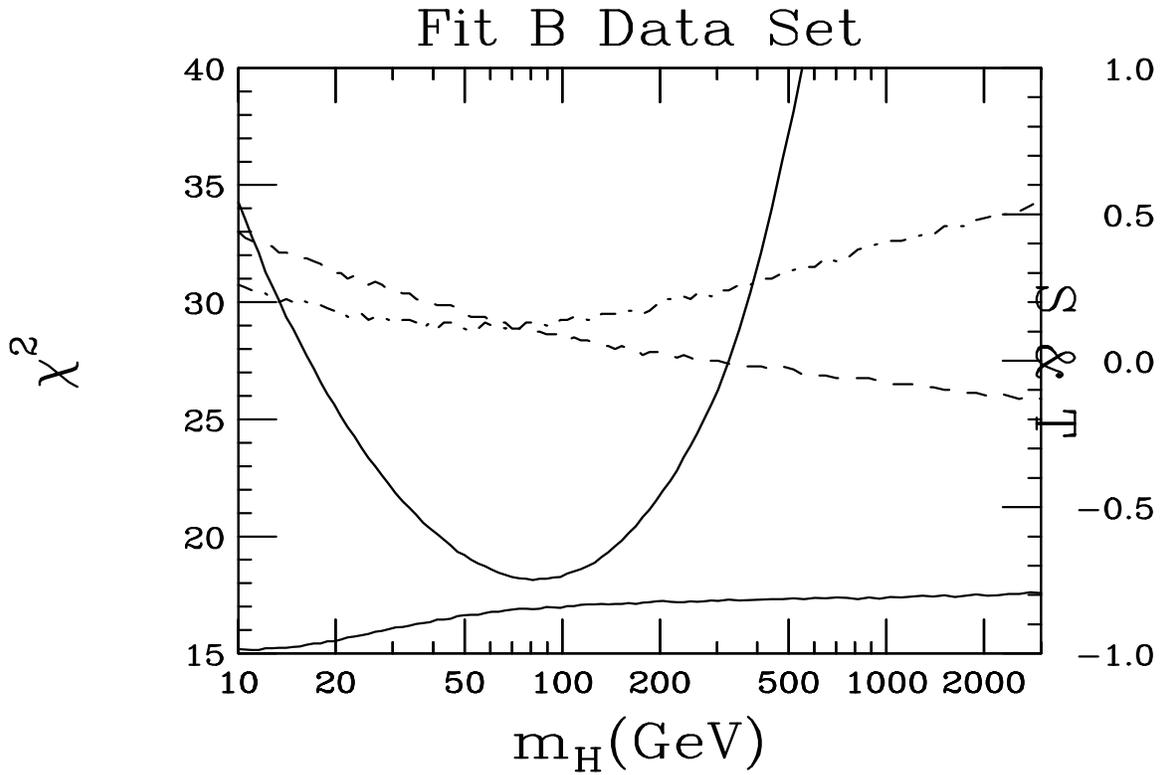}
\end{center}

\caption{$\chi^2$ distributions (solid lines) for the SM and $S,T$
fits to data set B, i.e., including the hadronic asymmetry
measurements but not $x_W^{\rm OS}[\stackrel {(-)}{\nu} N]$. $S$
(dashed line) and $T$ (dot-dashed line) are read to the scale on the
right axis.}

\label{fig12}
\end{figure}

\newpage
\begin{figure}

\begin{center}                                                                 
\includegraphics[height=6in,width=4in,angle=90]{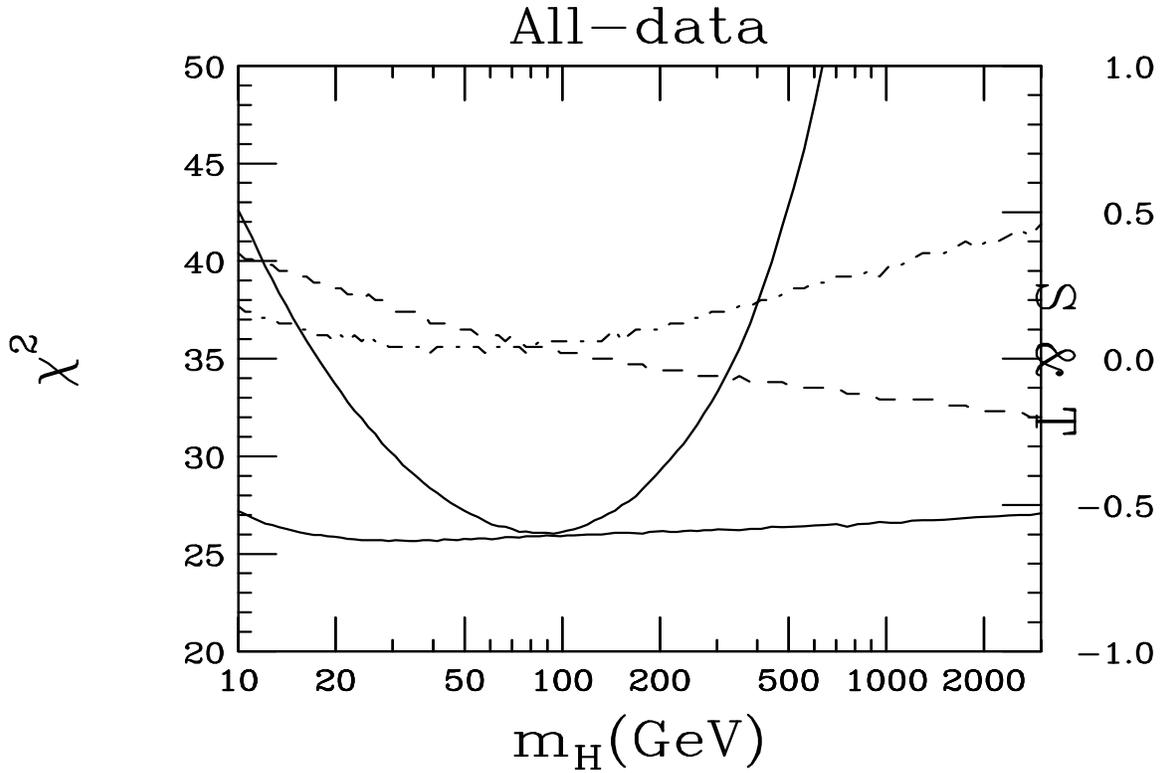}
\end{center}

\caption{$\chi^2$ distributions (solid lines) for $S,T$ fits to the
all-data set, data set A. The distribution for $S=T=0$ is not 
equivalent to the SM fit since it uses the model-independent NuTeV fit
to $g_{L,R}$ as discussed in the text. $S$ (dashed line) and $T$
(dot-dashed line) are read to the scale on the right axis. }

\label{fig13}
\end{figure}

\newpage
\begin{figure}
\caption{Electroweak schematic diagram.}

\begin{center}                                                                 
\includegraphics[height=4in,width=6.5in,angle=-90]{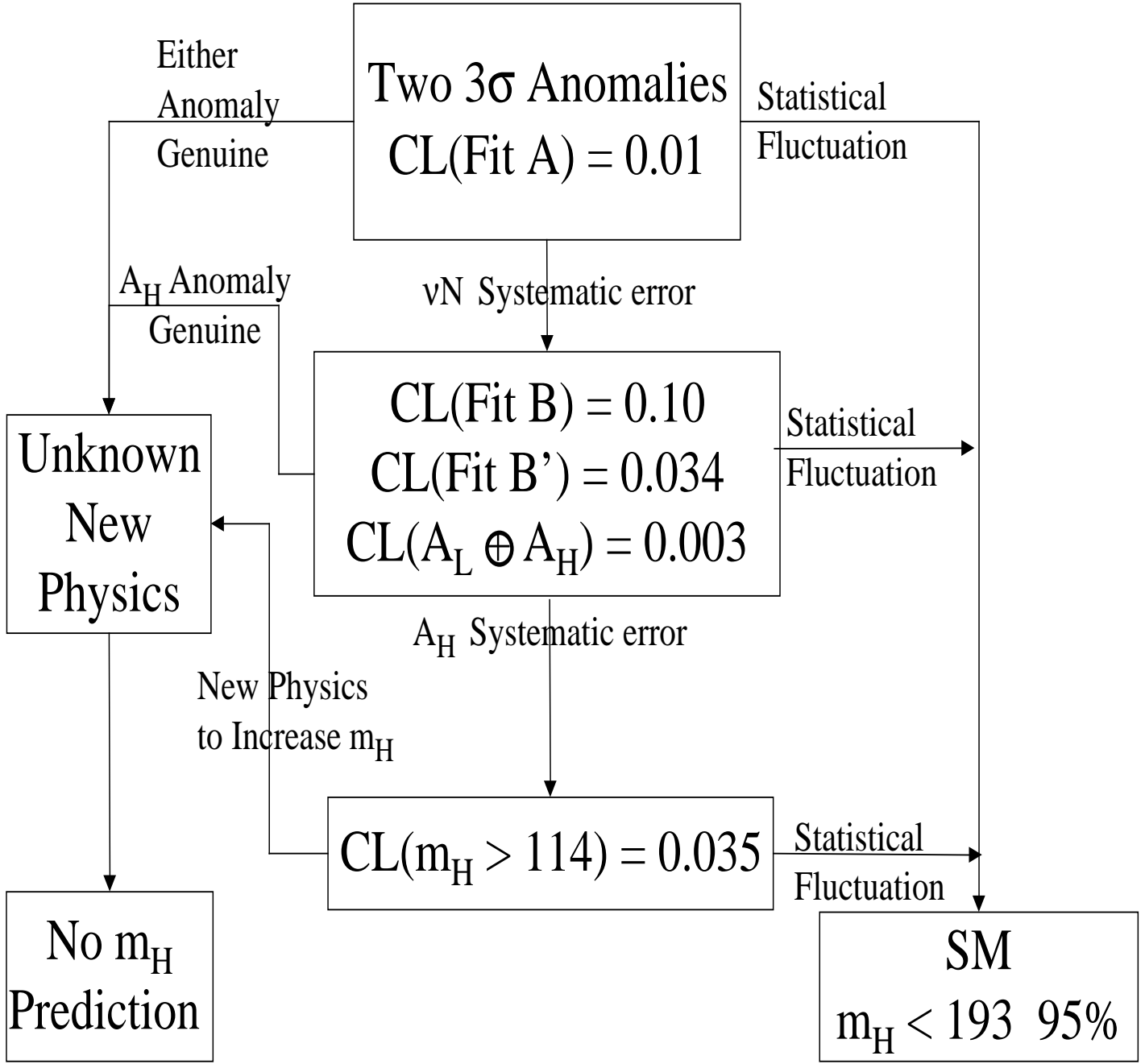}
\end{center}


\label{fig14}
\end{figure}

\end {document}